\documentclass[%
reprint,
 amsmath,
 amssymb,
 aps,
pre,
]{revtex4-2}

\usepackage{graphicx}
\usepackage{dcolumn}
\usepackage{bm}
\usepackage{hyperref}

\begin{document}

\preprint{}

\title{On the transformation of the Maxwell-Boltzmann Distribution to a Power-Law}

\author{Ari Laor}
    \email{laor@physics.technion.ac.il} 
\author{Igor Gitelman}          \email{giigor@campus.technion.ac.il}
\affiliation{Physics Department, Technion - Israel Institute of Technology, Haifa 32000, Israel 
}%

\date{\today}

\begin{abstract}

Power-law (PL) distribution functions (DF) are prevalent in highly diverse systems. 
The systems range in size from nanometer to mega light years, in complexity from dust grains
to living organisms, and characterize the distribution of various events in nature and in  
various human activities. To gain some insight on why PL DF are so prevalent, we explore the 
conditions leading to the formation of a PL DF in a simple system of colliding 
hard sphere. We follow the time evolution of the energy DF through direct Monte Carlo simulations.
In statistical equilibrium, the DF
evolves into the Maxwell-Boltzmann (MB) DF. A transition to a PL DF occurs when: 
1. The system is initially far from equilibrium. For example, a mix of light and
heavy particles with the same velocity. 2. The system dynamics is scale-free,
which holds in the intermediate asymptotic regime, far from the initial 
and the final equilibrium states. The scale-free  dynamics leads to a DF which evolves 
in a self-similar form. 3. The system is open with a scale-free  boundary condition. 
For example, a constant injection of particles far from equilibrium. 
The DF PL index is set by the time dependence of the self-similar DF and by the boundary
condition. The PL index is independent of the self-similar DF form.
Conditions 1-3 are common in a great variety of systems, which may explain why PL DF are so prevalent in nature.

\end{abstract}

\maketitle


\section{\label{sec:intro} Introduction}
Power law (PL) distribution functions (DF) are
prevalent in a great variety of systems. For example, the mass DF follows a PL
from the scale of dark matter halos, the most massive objects in the universe,
with masses of up to $\sim 10^{48}$~gr, down to the DF of dust grains in the interstellar medium,
with masses of $\sim 10^{-17}$~gr. A PL DF also characterizes the frequency  
of the energy release in various events, ranging in energy from cosmological gamma ray bursts, down to 
Solar eruptions and Earthquakes. PL DF also characterize the distribution of the country populations,
and the distribution of the population from small villages to largest cities
in a given country.
It also characterizes the DF of income (Pareto law), the frequency distribution
of the use of words (Zipf's law), and even the DF 
of mortality in wars and terrorist attacks \cite[e.g.][]{newman05, gabaix09}.  
Why are PL DF so prevalent in such highly diverse systems? 

Clearly, the formation of a PL DF is set by the system dynamics, and evolves given some specific initial and boundary conditions.
Indeed, some specific analytic approximations for the dynamics of some systems lead to
PL analytic solutions \cite[e.g.][]{sr06, aschwa16}.
However, the fact that PL DF are so prevalent in highly diverse systems 
suggests there are likely more general and prevalent 
conditions leading to the formation of a PL DF. 

The purpose of this work is to try and gain some insight on these general conditions, 
by simulating the time evolution of the DF of a simple physical system,
which can be followed to high precision with no approximations involved.

We note in passing that PL DF are sometimes confused with PL functional relations
between various quantities found in nature, such as the PL relation in living organisms between their body mass
and metabolic rate \cite{Kleiber47}. Similarly, PL DF should not be confused with PL scaling relations 
in various physical systems, such as the relation between the pendulum frequency and its length,
which can be derived from dimensional analysis (the so-called solutions of the first kind, 
\cite{Barenblatt96}).

Below we follow the formation of a PL DF in a simple physical system of colliding hard spheres.
The advantage of this system is its simple dynamics, which can be followed
accurately with no approximations, both in the classical and the relativistic regimes.
In contrast with more realistic systems, where some approximations are inevitable. 
In principle, the time evolution of the velocity DF, $f(v)$, of such a simple system 
can be followed by solving the Boltzmann equation for given initial and boundary condition. 
However, a direct solution of this multivariable integro-differential   
partial differential equation is notoriously difficult \cite[e.g.][]{mp06,wu13,ct15}, even for the 
simple case of colliding hard spheres. A full numerical solution of the
Boltzmann equation is similarly challenging. The equation is therefore often simplified, e.g. by
assuming the modeled system is only slightly offset from statistical equilibrium, or that the velocity changes per collision are small, so the collision integral can be replaced by a diffusion term (e.g. the Fokker-Planck equation, the Langevin equation).
The diffusion approximation can also be applied to a system of non-relativistic electrons + photons (Kompaneets equation \citep{RL79}). Another simplification is possible when the collision term can be neglected, leading to the collisionless Boltzmann equation. Some attempts were made to derive analytic PL solutions for the Boltzmann equation using various assumptions \cite{kats76, collier93, ht06}. 

Instead of attempting to solve the Boltzmann equation, we use direct Monte Carlo simulations to derive the time evolution 
of the DF of a system of colliding hard spheres, with initial conditions far from equilibrium. 
The Monte Carlo method is simple to implement, and allows a fast and accurate solution for the 
time evolution of $f(v)$, with no approximations involved. The only limitation is the finite system 
size it can handle, which is addressed by following the convergence of the results with the increasing number 
of particles in the system.

The paper is organized as follows, in section 2 we review the analytic solutions which lead to a PL
DF in a system where the dynamics is scale-free, and where the time dependent solution is self-similar.
In section 3 we describe the algorithm used for the direct Monte Carlo simulation in the non-relativistic and in the relativistic
regimes, the code validation scheme, and its application for various cases with initial conditions far from equilibrium.
The simulation results are described in section 4, discussed in section 5, and section 6 gives the main conclusions.

\section{\label{sec:analytic} Analytic estimates}
Below we review the analytic description of scale-free  dynamics, the resulting formation
of a self-similar DF, and the conditions when a self-similar DF evolves to a PL DF.
We then apply these considerations to the specific case of colliding hard spheres.

\subsection{\label{sec:scale} scale-free  dynamics}
A scale-free  function satisfies
\begin{eqnarray}
    f(ax)/f(x)=a^k,
\end{eqnarray}
for all values of $x$, where $a$ and $k$ are real constants. That is $f(x)$ is a homogeneous function of degree $k$.
For example, a PL function, 
\begin{eqnarray}
    f(x)=f_0x^\alpha,
\end{eqnarray}
is a scale-free function, since
\begin{eqnarray}
    f(ax)/f(x)=a^\alpha,
\end{eqnarray}
i.e. it is a homogeneous function of degree $\alpha$ \cite[e.g.][]{Hankey1972}.

Euler's homogeneous function theorem states 
that the reverse also holds. That is any scale-free function, i.e.
a homogeneous function of order $k$, is necessarily a PL function
with an index $k$. 
This holds since by definition a scale-free function satisfies
\begin{eqnarray}
  f(ax)=f(x)\times C(a)\  ,  
\end{eqnarray}
where the scaling factor $C(a)$ is some general function of $a$. 
Differentiating both sides w.r.t $a$, $d/da$, yields
\begin{eqnarray}
xf'(ax)=f(x)\times C'(a)\ ,
\end{eqnarray}
which gives for $a=1$ (without loss of generality)
\begin{eqnarray}
\frac{f'(x)}{f(x)}=\frac{C'(1)}{x},
\end{eqnarray}
that is
\begin{eqnarray}
\frac{d}{dx}\ln[f(x) ]=\frac{d}{dx}[\ln x^{C'(1)}] ,
\end{eqnarray}
with the solution 
\begin{eqnarray}
f(x)=Ax^{C'(1)} .
\end{eqnarray}
Thus, $f(x)$ is a PL function, with an index $\alpha=C'(1)$, 
and the general scaling factor $C(a)$ is necessarily 
of the form $a^k$, as $C(a)=f(ax)/f(x)=a^{C'(1)}$. To summarize, 
we have a one to one correspondence, a PL function is scale-free, and a scale-free  function is 
a PL.

If the physics describing a given process is scale-free  within a certain range of physical 
parameters, 
then the solution is also scale-free . However, a time dependent problem cannot
be scale-free, as the solution 
necessarily involves specific time dependent scales set by the initial values. 
However, as discussed below, if the system evolves far from the initial values, then
the effect of the initial values can be transient, and the dynamics becomes scale-free . The system 
solution then relaxes to a scale-free  solution, i.e. to a PL solution for a proper boundary condition.

\subsection{\label{sec:Self-sim} Self-similarity}
When the system dynamics is scale-free, the time evolution of the DF evolves
into a self-similar solution. 
Below we discuss the conditions when the self-similar solution evolves to a PL DF. 

Let us assume the DF represents the energy distribution, $dN/dE$, of a system of 
$N_p$ particles, injected into the system at a specific energy $E_{\rm inj}$. The system dynamics
leads to an evolution with time of $dN/dE$. If the DF becomes self-similar,
then by definition it can be written as
\begin{eqnarray}
\frac{dN}{dE}=g(t)f(E, E_0)\ ,
\end{eqnarray}
where $g(t)$ is a time dependent normalization factor of the DF, 
$E_0(t)$ is a time dependent characteristic energy of the system,
and $f(E, E_0)=f(x)$ is the self-similar DF, where $x=E/E_0$ . Note that a more general
functional form $x=g(E/E_0)$ is also viable. For example, $x=e^{E/E_0}$ (eq.~\ref{eq:transformation} below). 

The DF satisfies
\begin{eqnarray}
N_p=\int_0^\infty \frac{dN}{dE}dE .
\end{eqnarray}
Changing the integration variable from $E$ to $x$, which is dimensionless, gives
\begin{eqnarray}
N_p=g(t)E_0(t)\int_0^\infty f(x) dx .
\end{eqnarray}
The dimensionless explicit integral gives some number $C_1=\int_0^\infty f(x) dx$.
The DF normalization factor therefore satisfies 
\begin{eqnarray}
g(t)=N_p C_1^{-1}E_0(t)^{-1} .
\end{eqnarray}

We further assume continuous injection of particles at some rate,
$dN_p(t)/dt$, at a given specific energy $E_{\rm inj}$. 
Below we find the resulting time dependent DF, designated as $F(E,t)$.
We limit the derivation to an intermediate energy range $E_1$ - $E_2$, instead of $0$ - $\infty$, 
which satisfies $E_{\rm inj}\ll E_1$, and $E_2 \ll E_0(t)$, so $E$ at any moment is far from the 
two specific energies of the system, the initial energy and current DF characteristic energy, or the lowest 
and highest energies in the example studied below.
We further assume that by the time $t_1$, defined by $E_0(t_1)=E_1$, the DF relaxes from the initial
($t=0$) $\delta$ function at $E_{\rm inj}$ to the self-similar function $f(x)$.
The DF at a given $t$, $F(E,t)$, is derived by integrating over the contributions of the time evolving
DF of the particles injected starting from $t=0$,
\begin{eqnarray}
F(E,t)=\int_{0}^{t} \frac{dg}{dt} f(E,E_0(t')) dt'\ .
\end{eqnarray}
Changing the integration variable from $t'$ to $E_0$, and confining the
integration to $E_1$ - $E_2$, i.e. the energy range where the dynamics is scale-free, gives
\begin{eqnarray}
F(E,t)=\int_{E_1}^{E_2} \frac{dg}{dt'} f(E,E_0(t')) \frac{dE_0}{dE_0/dt'}.
\end{eqnarray}
We now substitute $E_0=E/x$, and $dE_0=-E\times dx/x^2$, which gives
\begin{eqnarray}
F(E,t)=E \int_{x_1}^{x_2}\frac{dg/dt'}{dE_0/dt'} f(x) x^{-2}dx.
\end{eqnarray}
The exact functional form of the self-similar function $f(x)$ is not relevant, 
as the specific integral approaches some constant value, $C_1$, when the boundaries satisfy 
 $x_1\ll 1$ and $x_2\gg 1$. 

\subsection{\label{sec:DerivGamma} A PL solution}

When do we expect to get a PL DF solution, i.e. $F(E)\propto E^{\Gamma}$?
An obvious guess is for the case of a PL dependence on time for
both $N_p(t)\propto t^a$ and $E_0(t)\propto t^b$. These relations imply
\begin{eqnarray}
\frac{dg/dt'}{dE_0/dt'} \propto t^{a-2b}\propto E_0^\frac{a-2b}{b}
\end{eqnarray}
Substituting in the above integral then gives
\begin{eqnarray}
F(E,t)\propto E^{\frac{a}{b}-1} \int_{x_1}^{x_2} f(x) x^{-\frac{a}{b}}dx.
\end{eqnarray}
Thus, $F(E)$ forms a PL with an index 
\begin{eqnarray}
\Gamma=a/b-1\ .  \label{eq:index}
\end{eqnarray}
For example, in the case of a constant injection rate, i.e.
$a=1$, and second order Fermi acceleration of non-relativistic particles, 
i.e. $b=2$ (see below), gives $F(E,t)\propto E^{-1/2}$.

The other functional forms which lead to a PL solution are
exponential dependencies,
$N_p(t)\propto e^{\alpha t}$ and $E_0(t)\propto e^{\beta t}$.
Repeating the derivation above leads to a similar PL solution for
$F(E,t)$, with 
\begin{eqnarray}
\Gamma=\alpha/\beta -1\ . \label{eq:index.greek}
\end{eqnarray}

Although exponential functions are not scale-free, but rather depend
on the characteristic scales, the times $1/\alpha$ and $1/\beta$ in the example above, 
the ratio of the characteristic times is dimensionless, and leads to the PL 
relation, $N_p\propto E_0^{\alpha/\beta}$. The relation between
$N_p$ and $E_0$ is therefore scale-free, although the dependence of each on $t$
is not scale-free . The PL relation between $N_p$ and $E_0$ is
similar to the one derived assuming PL dependence for each, 
which gives $N_p\propto E_0^{a/b}$.

What happens in a mixed state, say an exponential evolution of $E_0(t)$
and a PL evolution of $N_p(t)$? Let us explore the specific case of a 
constant injection rate, $\dot{n}$,
i.e. $N_p=\dot{nt}$, and an exponential energy change 
$E_0(t)=E_1 e^{\beta t}$. These relations give
\begin{eqnarray}
\frac{dg/dt'}{dE_0/dt'} =\frac{dN_p/dt}{E_0 dE_0/dt}=\frac{\dot{n}}{b} \frac{1}{E_0^2},
\end{eqnarray}
and therefore 
\begin{eqnarray}
F(E,t)\propto E^{-1} \int_{x_1}^{x_2} f(x) dx , \label{eq:mixed-state}
\end{eqnarray}
that is an index of $-1$. 

As one can show, a mixed state with $a \ne 1$,
or the general case of an exponential evolution of $N_p(t)$ and a PL evolution of $E_0(t)$,
does not lead to a PL solution for $F(E,t)$. Thus, a system 
characterized by a self-similar DF, $f(x)$, does not necessarily evolve  into
a scale-free  DF. To get a scale-free, i.e. a PL DF,  $N_p(t)$ and $E_0(t)$
should both have a PL dependence on $t$, or both should evolve exponentially with $t$.
A linear evolution of $N_p(t)$, together with an exponential evolution of $E_0(t)$,
also leads to a PL DF. 

We stress that the PL solution applies only for the
fraction of the particle population which reside at the intermediate energy regime.

\subsection{\label{sec:Application} Application for colliding hard spheres}
Below we apply the above analytic derivation for the specific case
of colliding hard spheres, which is far from statistical equilibrium.

We assume the hard spheres are effectively point like particles, and that the particles
are either light or heavy. 
Initially, we calculate the energy exchange per collision between a light particle and a heavy one.
The derivation holds for both non-relativistic $\gamma\beta\ll 1$
kinematics and relativistic $\gamma\beta\gg 1$ kinematics, where $\gamma$ and $\beta$ are
the Lorentz factor and the ratio of the particle velocity to the speed of light.
To simplify the analytic derivation below we make the following assumptions: 1) 
the light and heavy particles masses satisfy $m_l\ll m_h$, 2) the particles 
are reflected backward in the center-of-mass frame. These assumptions lead to 
\begin{eqnarray}
\frac{\Delta E_l}{E_l}=2\gamma_h^2 \left( \beta_h^2-(\vec{\beta}_h\cdot \vec{\beta}_l)\right),
\end{eqnarray}
where  $E_l,\ \Delta E_l$ and $\vec{\beta}_l$ are the initial energy, energy change, and 
velocity divided by the speed of light (hereafter $c=1$) of the light particle, while $\gamma_h$ and $\vec{\beta}_h$ are the Lorentz factor and velocity for the heavy particle. The mean over incident angles gives 
\begin{eqnarray}
\frac{\Delta E_l}{E_l}=2\gamma_h^2\beta_h^2 \label{eq:rel_energy_gain}
\end{eqnarray}
Generally $E_l=m_l+E_{kl}$, where $E_{kl}$ is the kinetic energy of the light particle.
In the non-relativistic limit, $E_{kl}\simeq \frac{1}{2}\times m_l v_l^2$, 
where $E_{kl}\ll m_l$,
$v_l, v_h$ are the velocity of the light and heavy particle, and $\gamma_h \simeq 1$, which gives
a mean fractional kinetic energy gain for the light particle of
\begin{eqnarray}
\frac{\Delta E_{kl}}{E_{kl}}\simeq 4 \frac{v_h^2}{v_l^2}, ~~~ {\rm or}~~ \Delta E_{kl}\simeq 2m_l v_h^2. \label{eq:non_rel_energy_gain}
\end{eqnarray}
That is, a constant $\Delta E_{kl}$ per scattering. The energy growth is therefore linear with the 
number of scattering. In contrast, in the relativistic limit (eq.~\ref{eq:rel_energy_gain}), the relative 
energy gain and not the absolute energy gain is constant, leading to an exponential energy growth with the number of scattering. 
Note that in both cases the heavy particle is non-relativistic. The only assumption is that it is massive enough
that its kinetic energy can be assumed to remain constant, that is  $\Delta E_{kl}/E_{kh}\ll 1$.

In the numerical simulations below the light and heavy particles start with the same initial non-relativistic velocity. We therefore expect the mean energy of the light particles to initially grow linearly with the number of scattering, $N_c(t)$.
That is
\begin{eqnarray}
E_0(t)=C_2 N_c(t)+ E_{\rm inj}\ , 
\end{eqnarray}
where $E_{\rm inj}$ is the initial energy of the light particles, and $C_2$ is a constant
set by the particles velocity. 
Self-similarity is expected when $E_0(t)\gg E_{\rm inj}$,
as then $E_0(t)\simeq C_2 N_c(t)$. When 
the light particles become relativistic, their characteristic energy scales
exponentially with time (eq.~\ref{eq:rel_energy_gain}), where
\begin{eqnarray}
E_0(t)=E_{\rm inj}\times  e^{2\gamma_h^2\beta_h^2 N_c(t)} , \label{eq:exponential}
\end{eqnarray}
leading to exponential self-similarity, instead of the multiplicative self-similarity
of the non-relativistic regime. Note that if the heavy particles are also relativistic ($\gamma_h\gg 1$), the 
light particles characteristic energy growth time becomes exponentially short, and near equipartition
will be achieved following just a single scattering.

The solution for $N_c(t)$ is given by the collision rate of a light particle with the heavy ones, which is
\begin{eqnarray}
f_c\equiv \frac{dN_c}{dt}=\rho\sigma v , \label{eq:f_c}
\end{eqnarray}
where $\rho$ is the number density of the massive particles 
and $\sigma$ is the collision cross section, both assumed constant
for simplicity, and $v$ is their characteristic relative collision velocity.
Since the light particles kinetic energy increases with time, in contrast with
the massive particles which gradually loose energy and slow down, then by the time that $E_0(t)\gg E_{\rm inj}$ we get $v_l\gg v_h$, and thus $v\simeq v_l$.
For the non-relativistic case, $v=\sqrt{2E_0 m_l}=\sqrt{2E_0}$, for $m_l=1$. 
The equations above give,
\begin{eqnarray}
\frac{d}{dt} C_2^{-1}E_0 = \rho\sigma \sqrt{2E_0}\ ,
\end{eqnarray}
or
\begin{eqnarray}
\frac{dE_0}{dt} = C_2\rho\sigma\sqrt{E_0}\ ,\ {\rm i.e.}\ \ E_0^{-1/2}dE_0 = C_2\rho\sigma dt\ ,
\end{eqnarray}
with a solution, 
\begin{eqnarray}
E_0(t)\propto t^2\ ,\ {\rm and\ therefore}\ \ N_c(t)\propto t^2. \label{eq:E_0t^2}
\end{eqnarray}
In the ultra-relativistic limit, 
the collision rate $\rightarrow \rho\sigma c$, i.e. approaches a
constant, and therefore $N_c(t)\propto t$.

In the more general case of an energy dependent cross section, where
\begin{eqnarray}
\sigma(E_0) \propto E_0^{\delta}\ ,  \label{eq:sigmagamma}
\end{eqnarray}
the derivation above yields
a time evolution of $E_0(t)$ which scales as
\begin{eqnarray}
E_0(t)\propto t^{\frac{2}{1-2\delta}}\ ,   \label{eq:indexgamma}
\end{eqnarray}
i.e. $b=2/(1-2\delta)$, which implies (eq.~\ref{eq:index})
\begin{eqnarray}
\Gamma=a/2-a\delta-1\ .  \label{eq:index.delta}
\end{eqnarray}

The cumulative DF at a given $t$ for the constant injection rate $\dot{n}$ is given by
\begin{eqnarray}
F(E,t)=\dot{n}\int_0^t f(E,E_0(t')) dt'
\end{eqnarray}
Based on the simulation below, we use $f(x)=x^\alpha e^{-x}$ ($x=E/E_0$), for the self-similar time evolution DF of particles injected at $t'$ to $t'+dt'$, where $N_p=\dot{n}dt'$. 
As shown above, the value of the DF PL index, $\Gamma$, is independent of the specific functional form
of $f(x)$, since $f(x)$ is integrated out.

We change the integration
variables from $t$ to $E_0$, using the relation above $dt= C_3^{-1}E_0^{-1/2}dE_0$, which gives
\begin{eqnarray}
F(E,t)=\int_0^t \dot{n} C_1^{-1}E_0^{-1} (E/E_0)^{\alpha}e^{-E/E_0} dt\ ,
\end{eqnarray}
and
\begin{eqnarray}
F(E,t)= && \,\dot{n} C_1^{-1}C_3^{-1} E^{\alpha} \nonumber \\
 && \times  \int_{E_0(0)}^{E_0(t)} E_0^{-1} 
E_0^{-\alpha}e^{-E/E_0}E_0^{-1/2}dE_0 \ ,
\end{eqnarray}
or
\begin{eqnarray}
F(E,t)=&&\,\dot{n} C_1^{-1}C_3^{-1}  E^{\alpha}\nonumber \\  &&\times\int_{E_0(0)}^{E_0(t)}  E_0^{-\alpha-3/2}e^{-E/E_0}dE_0 .
\end{eqnarray}
Substituting $E_0=E/x$ and $dE_0=-E\times dx/x^2 $, and integrate over $x$
\begin{eqnarray}
F(E,t)= &&\, -\dot{n} C_1^{-1}C_3^{-1} E^{\alpha}\nonumber \\  \times && \int_{x(0)}^{x(t)} (Ex)^{-\alpha-3/2}e^{-E/E_0}E/x^2 dx
\end{eqnarray}
which gives
\begin{eqnarray}
F(E,t)=&&\,-\dot{n} C_1^{-1}C_3^{-1}E^{-1/2}\nonumber \\ &&\times  \int_{x(0)}^{x(t)}  x^{-\alpha-7/2}e^{-x} dx
\end{eqnarray}
The integral should be roughly constant since  $x(0)\rightarrow \infty$ and $x(t)\ll 1$,
so the integral converges to a definite integral at $0 - \infty$.
In this case we get 
\begin{eqnarray}
F(E) \propto E^{-1/2}\ . \label{eq:f0.5}
\end{eqnarray}
This PL scaling of $F(E)$ with an index $\Gamma=-1/2$, matches the value derived above
for the general case of a self-similar evolution (eq.~\ref{eq:index}), with $a=1$ and $b=2$.
One can show that a similar steady state analytic solution in the relativistic 
limit leads to $F(E) \propto E^{-1}$ (eq.~\ref{eq:mixed-state}).

Below we follow numerically the time evolution of the DF of various systems far from
equilibrium. The evolution is followed using Monte Carlo direct simulations 
which solves explicitly the binary collisions in the system. The analytic
steady state solutions above are used below as one of the code validation tests.

\section{\label{sec:sim} The Simulation algorithm}
\subsection{\label{chap:outlines} General outline}
The numerical simulations effectively provides a solution for the Boltzmann equation, i.e. the 
time evolution of the velocity DF, $f(v,t)$, for a system of $N_p$ colliding hard spheres. We use the Direct Simulation Monte Carlo (DSMC) method, which is a computational stochastic method used for solving problems involving the dynamics of dilute gas, where the fluid approximation breaks down and the dynamics of the individual particle needs to be followed (e.g. as done in 
molecular dynamics and particle in cell simulations).
We simulate systems which are spatially uniform, so $f(v)$ is independent of position, which 
greatly reduces the computation time.  The systems consists of multiple types of point like particles, each type characterized by their mass and initial velocity or kinetic energy. The initial angular distribution of the velocity is assumed to be isotropic, that is the simulation is at the global system rest frame. Given the absence of external forces
in our simulations, the overall isotropy of the velocity distribution is maintained.

The system evolves by following the collisions between the particles. 
We include only binary collisions, consistent with our assumption of point like hard sphere collisions, where the collision
time is effectively zero. The collisions are elastic so both energy and 
momentum are conserved. The probability of a given collision per unit time, in the non-relativistic limit, scales linearly with the relative speed of the colliding particles. 
Therefore, to determine if two particles collide in a given time step, we compare their collision rate, $f_c$, to the maximum rate found thus far, denoted as $f_{max}$, and a randomly generated number $\mathcal{R}$, which is uniformly distributed in the range 0 - 1. A collision is therefore considered to occur if  
\begin{eqnarray}
f>f_{max}\cdot\mathcal{R}\ . \label{eq:coll_rate_cond}
\end{eqnarray}

The system evolution is followed in time steps that correspond to a single collision for the two particles with the
highest relative velocity at a given time. That is 
\begin{eqnarray}
\Delta t= \frac{2}{N_p\cdot f_{max}}\ , \label{eq:time_incriment}
\end{eqnarray}
where the division is by $N_p/2$ since only one pair is selected and followed out of the maximal possible number
of $N_p/2$ pairs in each time step. The simulation time $t$ is a sum of all the time increments $\Delta t$. 

We initialize the simulation by assigning each particle $i$ (= 1 - $N_p$) with a velocity with the same amplitudes $v$, in 
the direction $\phi_i$ and $\theta_i$, where $\phi_i =2\pi \cdot\mathcal{R}$, and $\cos \theta_i = 1-2\cdot\mathcal{R}$.
To facilitate a collision, we randomly select two particles and calculate their collision rate, to determine if that collision occurs within the simulated $\Delta t$. The collision rate is calculated using 
\begin{eqnarray}
f_c = F \sigma \sqrt{(\vec{\beta}_1-\vec{\beta}_2)^2-[\vec{\beta}_1\times \vec{\beta}_2]^2} \label{eq:R-freq}
\end{eqnarray}
where $\vec{\beta}_1,\vec{\beta}_2$ are the particle velocity relative to the speed of light $c$ ($c=1$), $\sigma$ - is the collision cross section, and $F$ - a normalization constant (see the discussion in Appendix~\ref{apen:rel_freq}).

The collision of the two randomly selected particles with energy-momentum $(E_{1}, \vec{p}_{1}), (E_{2}, \vec{p}_{2})$, is most simply described in their center of mass (CM) frame. The CM velocity $\vec{\beta}$ is: 
\begin{eqnarray}
\vec{\beta}=\frac{\vec{p}_{1}+\vec{p}_{2}}{E_{1}+E_{2}}.
\end{eqnarray}
We follow the collision in the CM frame, where the total momentum remains zero. In addition, the total energy is conserved (hard sphere), 
and thus the individual momenta amplitudes and energies remain unchanged, but the momenta directions are changed randomly to new 
and opposite directions. We now transform back from the CM frame to the lab frame, and get the revised values of
$E$, and $\vec{p}$ for the two particles. The simulation above repeats by making a new random selection of two particles
out of the $N_p$ particles. The simulation time is advanced again by $\Delta t$. We then derive histograms of $f(v,t)$, and $f(E,t)$,
at certain $t$ values, to follow the time evolution of the DF of the light and heavy particles.

For the sake of completeness, the Lorentz transformations to a system with $\vec{\beta}$ at some arbitrary direction is given by (\cite{cushing1967LorentzVector})
\begin{eqnarray}
    &&E_{\rm f}=\gamma \left(E_{\rm in}-\vec{\beta}\cdot\vec{p}_{\rm in}\right)\nonumber 
    \\ &&\vec{p}_{\rm f}= \vec{p}_{\rm in}+(\gamma -1)(\vec{p}_{\rm in}\cdot\hat{n})\hat{n}-\gamma E_{\rm in} \vec{\beta} \label{eq:Lorentz_trans}
\end{eqnarray}
where  $\hat{n} = \vec{\beta}/\beta$, $\gamma = (1-\vec{\beta}^2)^{-1/2}$, and $E_{\rm in},\vec{p}_{\rm in}, E_{\rm f},\vec{p}_{\rm f}$ are the initial and final energy and momentum, that is before and after the transformation.

\subsection{\label{subsec:valid}The algorithm validation}

\subsubsection{\label{subsubsec:valid1} Identical particles}

We validate the code by comparing the convergence of the numerically derived $f(v,t)$, or equivalently  $f(E,t)$, on a long enough timescale with the statistical equilibrium analytic solutions. The general covariant statistical equilibrium solution for colliding hard spheres in a homogeneous
medium is the Maxwell–Juttner (MJ) distribution 
\begin{eqnarray}
F(E_k)=&&\frac{1}{N}\frac{dN}{dE_k}=\frac{1}{m^2 T \mathcal{K}_2^s(\frac{m}{T})}(E_k+m) \nonumber\\
& & \times \sqrt{(E_k+m)^2-m^2}\exp\left({-\frac{E_k}{T}}\right) \label{eq:MJ_Ke}
\end{eqnarray}
where $m$ is the particles mass, $T$ is the distribution temperature  (in energy units, i.e. using for convenience a Boltzmann coefficient
$k_{\rm B}=1$), $E$ the particle energy, where 
$E_k=E-m$ is the particle kinetic energy,  and  $\mathcal{K}_2^s(z)$ is the scaled Bessel function, which is 
related to the Modified Bessel Function of the Second Kind $\mathcal{K}_2(z)$ through $\mathcal{K}_2^s(z) = \mathcal{K}_2(z) \exp(z)$ \cite{Gradsh1965Tables}.

The equilibrium DF in the non-relativistic limit, where $E_k\ll m$, or equivalently $\gamma\beta\ll 1$, reduces to the Maxwell-Boltzmann (MB) DF 
\begin{eqnarray}
    F(E_k)=\frac{2}{\sqrt{\pi T^3}}\sqrt{E_k}\exp\left(-\frac{E_k}{T}\right).
\end{eqnarray}
The temperature is related to the mean particle energy, $\langle E \rangle = \frac{E_t}{N_p}$, where $E_t$ is the total particle energy, through the relation
\begin{eqnarray}
    \langle E \rangle = 3 T + m\frac{\mathcal{K}_1\left(\frac{m}{T}\right)}{\mathcal{K}_2\left(\frac{m}{T}\right)}.
    \label{eq:MeanEnergy}
\end{eqnarray}
In the non-relativistic limit $\frac{\mathcal{K}_1(x)}{\mathcal{K}_2(x)}\approx 1-\frac{3}{2x}$, for $x=m/T\gg 1$,
which gives $\langle E_k \rangle \approx \frac{3}{2} T$. In the ultra-relativistic limit, $x=m/T\ll 1$, $\frac{\mathcal{K}_1(x)}{\mathcal{K}_2(x)} \approx \frac{x}{2}$ , which gives $\langle E_k \rangle \approx 3 T$.

\begin{figure}[b]
\includegraphics[width=8 cm]{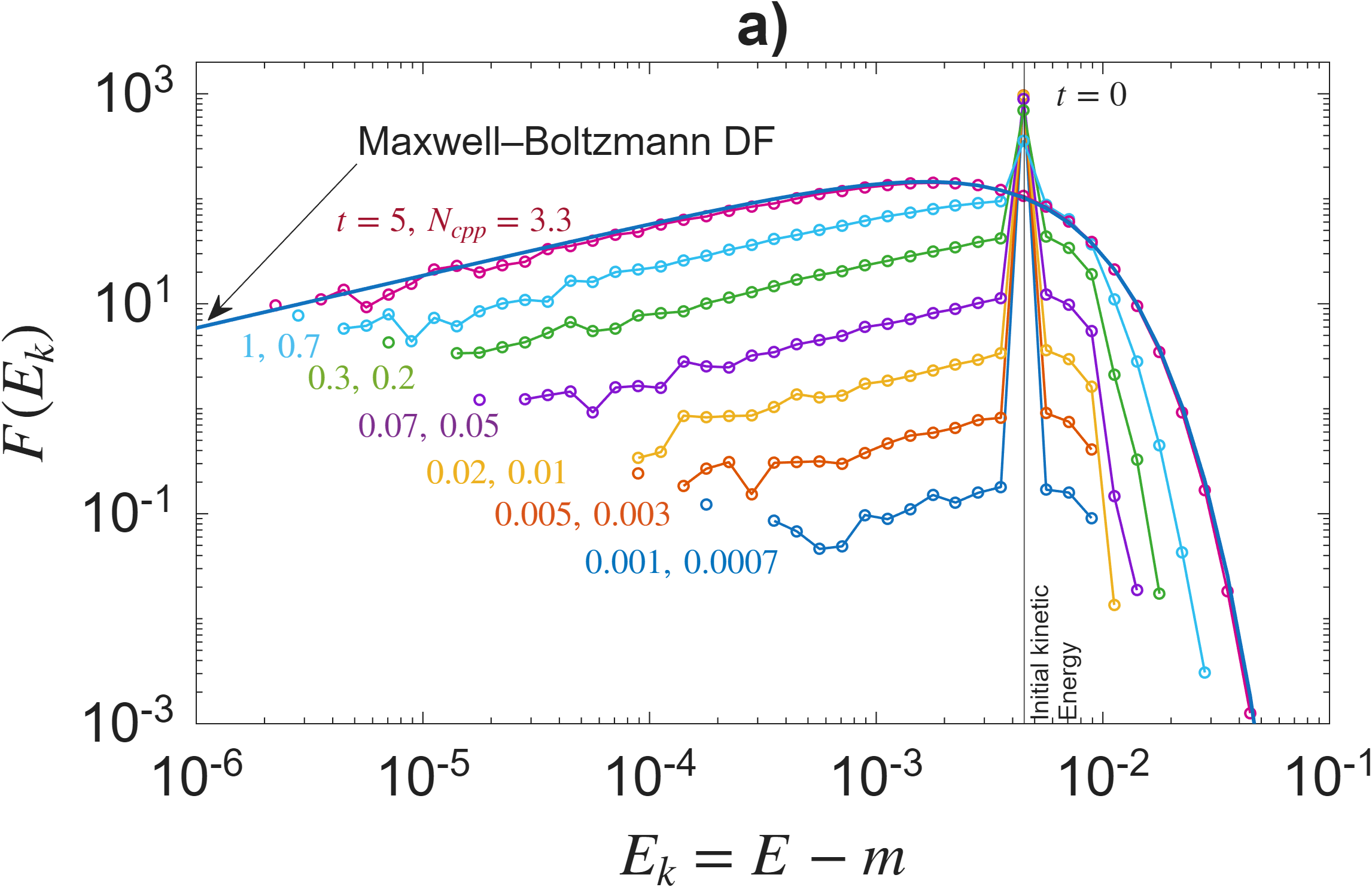}
\includegraphics[width=8 cm]{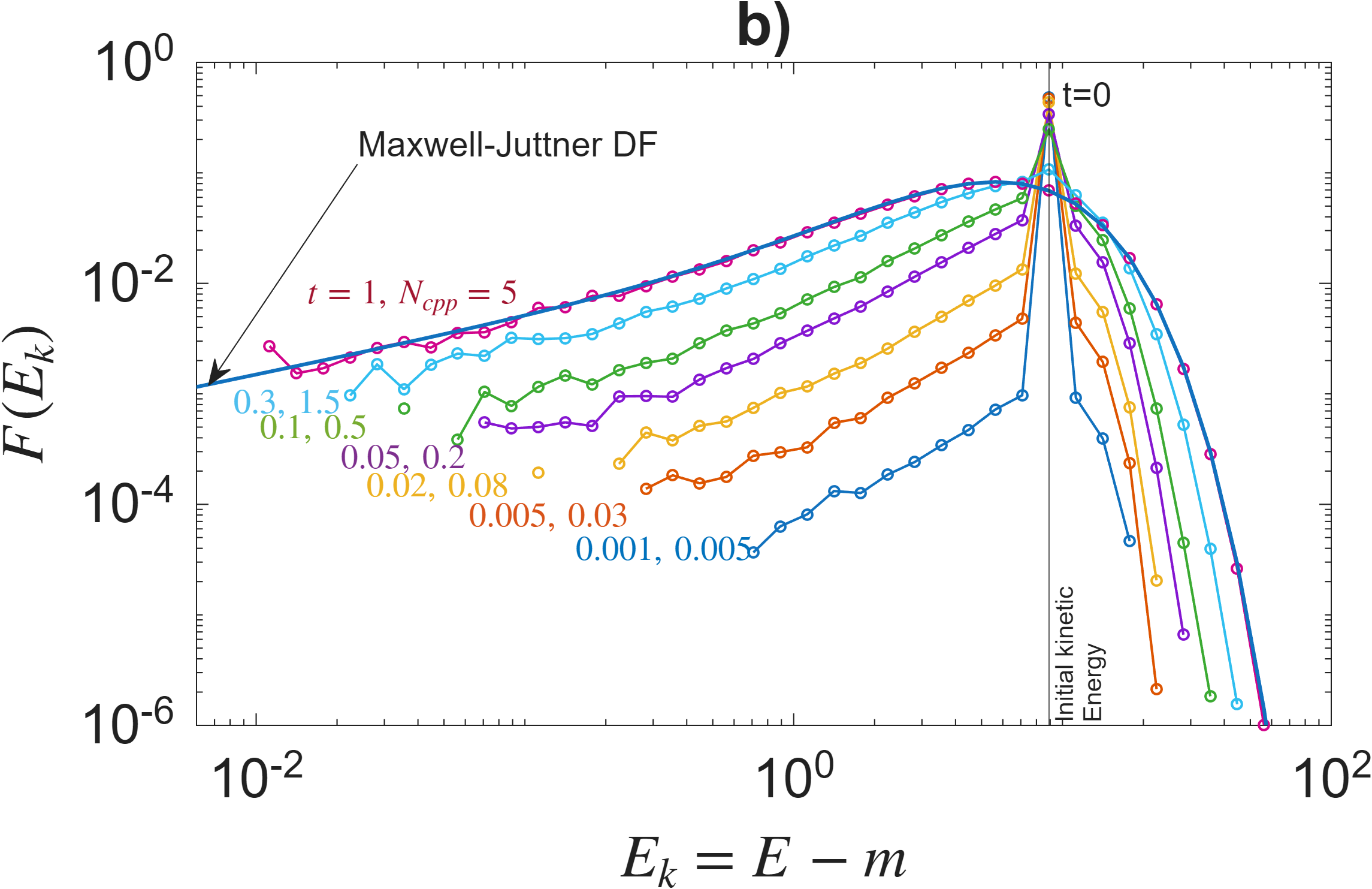}
\caption{\label{fig:code_val} The evolution of the DF of $E_k$ for a system of $10^6$ identical particles, starting from 
a $\delta$ function DF. \textbf{Panel a)} Non-relativistic particles with an initial DF of $\gamma \beta= 10^{-1}$. The DF 
evolves towards the equilibrium MB DF. The time step and the
mean number of collisions per particle $N_{\rm cpp}$ is noted near each curve. At $t=5$ ($t$ defined in 
eq.~\ref{eq:time_incriment}), where
$N_{\rm cpp}=3.3$, the simulation converges well to the MB solution.
\textbf{Panel b)} Relativistic particles with an initial $\gamma \beta= 10$ DF. The simulation converges to the MJ DF already at $t=1$, as $N_{\rm cpp}=5$. The high energy 
exponential solutions for MB and MJ are similar, but at low energies $F(E_k)\propto E_k^{1/2}$
in MB, while $F(E_k)\propto E_k^2$ in MJ. Note that the MJ DF also 
switches from $E_k^2$ to $E_k^{1/2}$ at $E_k < 0.1$, where the particles become non-relativistic.} 
\end{figure}

\begin{figure*}
\includegraphics[width=8 cm]{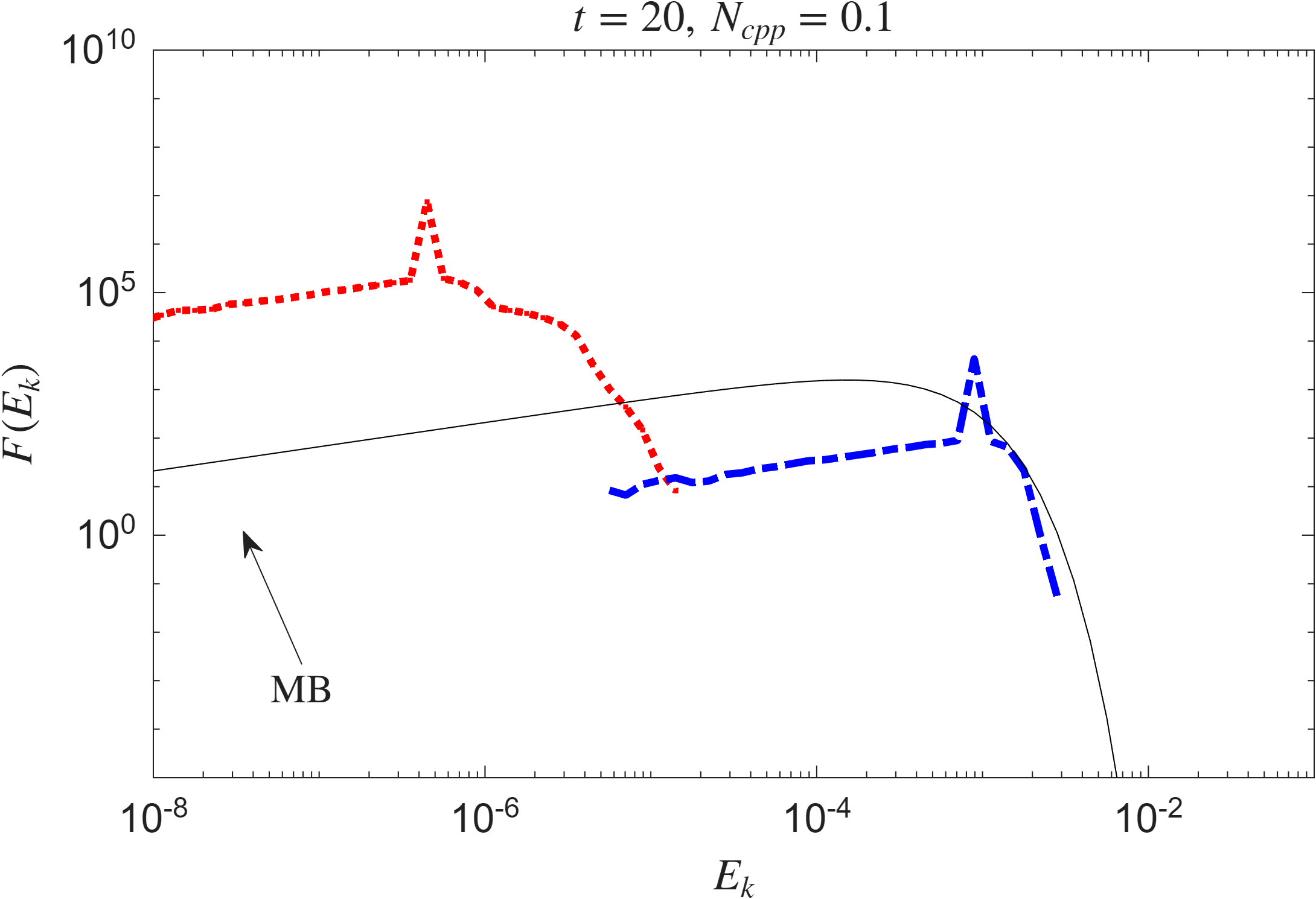}
\includegraphics[width=8 cm]{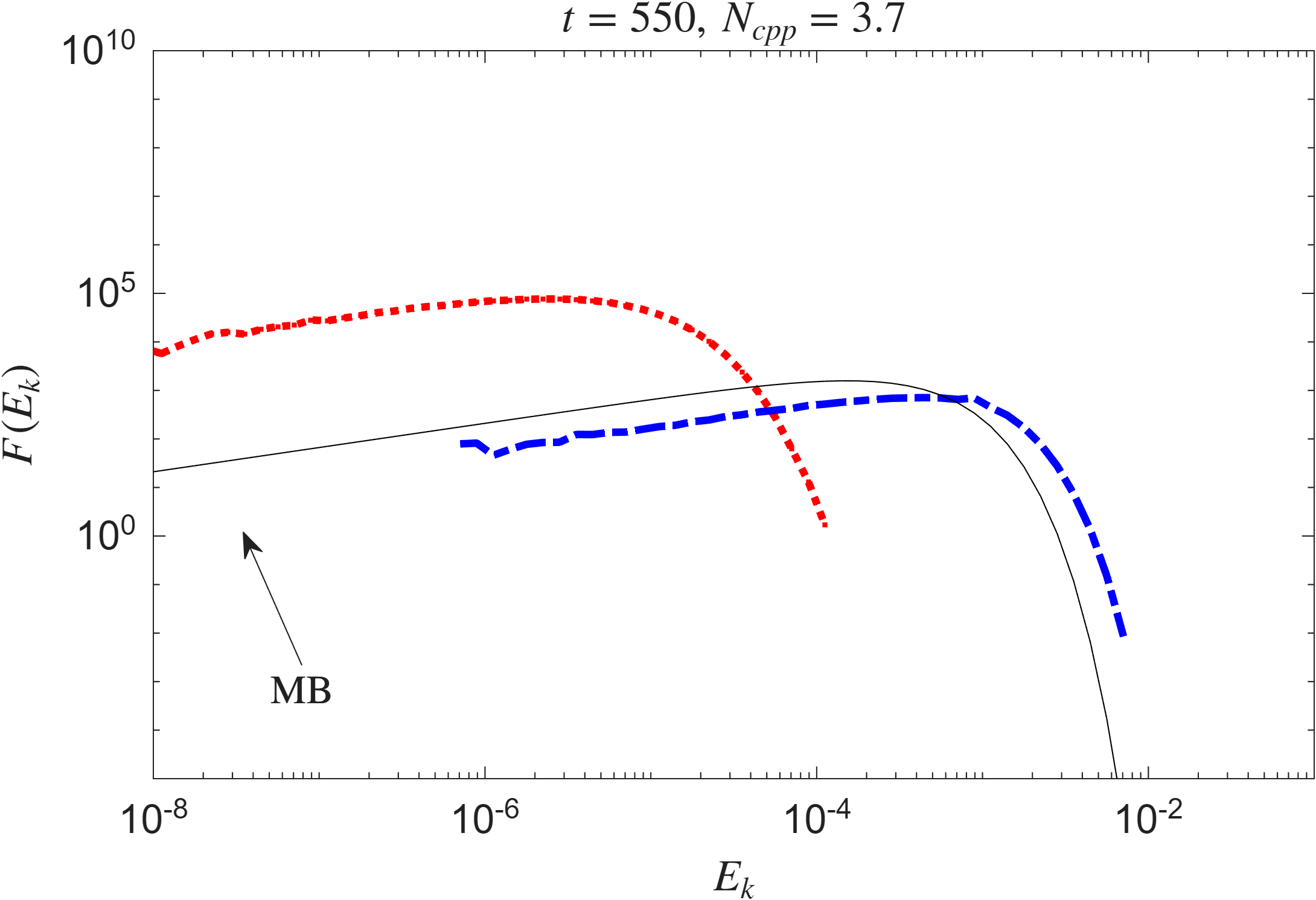}
\includegraphics[width=8 cm]{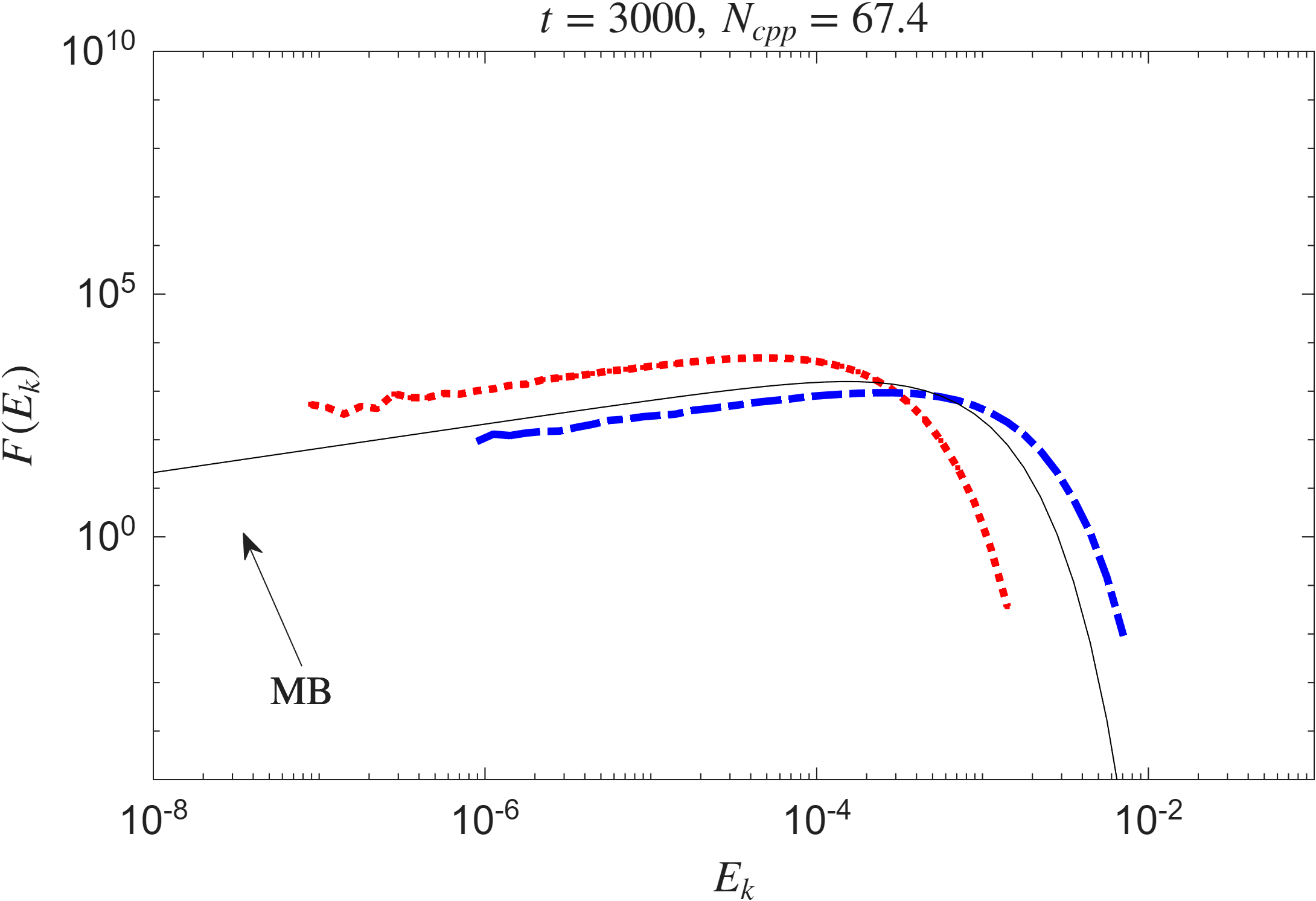}
\includegraphics[width=8 cm]{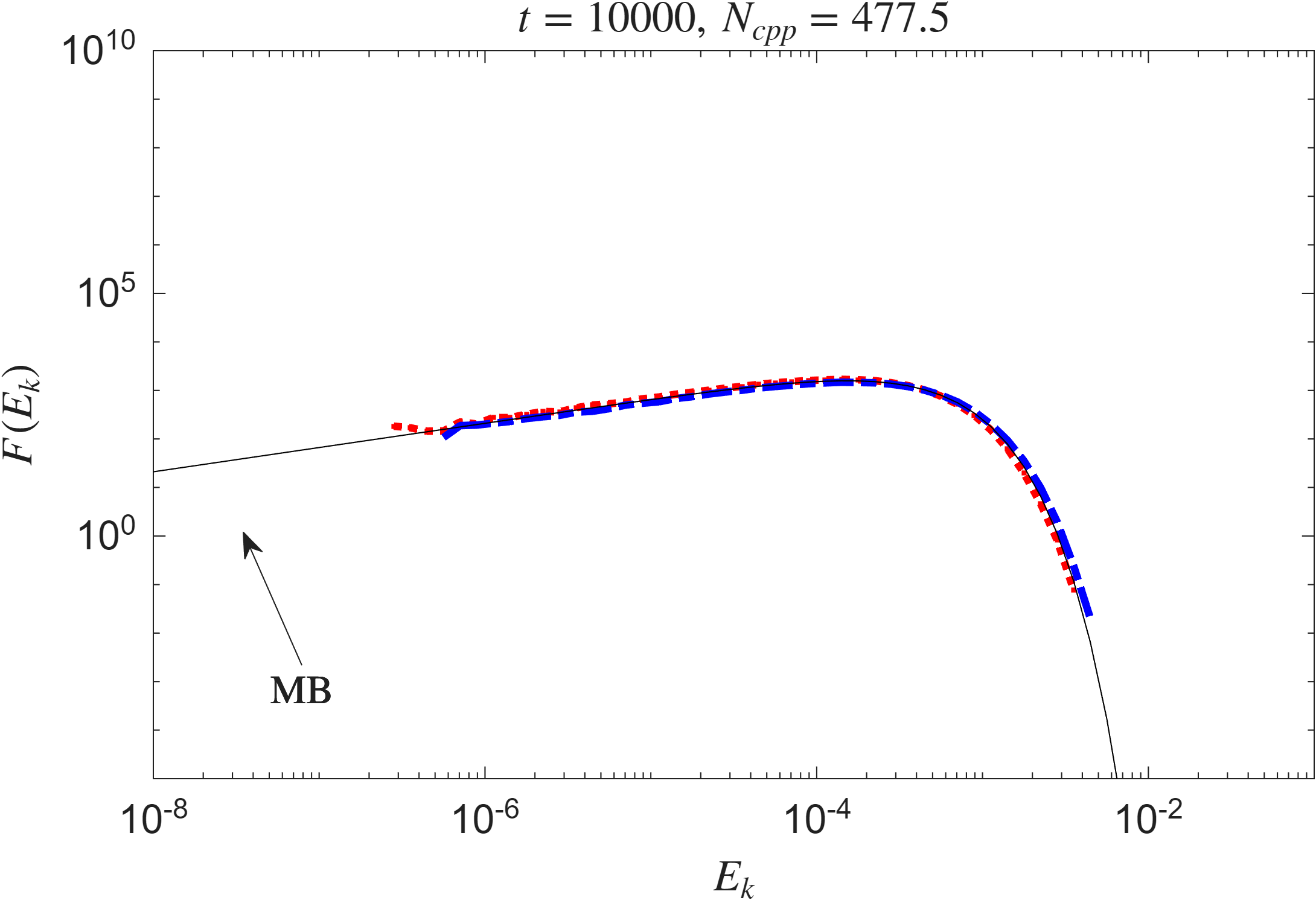}

\caption{\label{fig:code_val_two_type} The time evolution of the DF for a system of two types of particles,
low mass particles with $m_1=1$ (marked in red), and massive particles with $m_2=1836$ (marked in blue). The thin grey lines marks the final equilibrium DFs. The two populations start with highly discrepant $\langle E_k\rangle$, by a factor of
$m_2/m_1=1836$. Only at $N_{\rm cpp}=480$, reached at $t=1e4$, the system reaches a statistical equilibrium in energy space, i.e. equipartition with identical MB energy DF.} 
\end{figure*}

To validate the code we simulate a system of $N_p=10^6$ identical particles, all with the same mass and velocity magnitude
but at random directions. The particles collide according to the algorithm described in the previous subsection (\ref{chap:outlines}). We follow the time evolution of $f(E_k,t)$ until it reaches a statistically steady state solution, i.e. a statistical equilibrium,
and compare the numerically derived $F(E_k)$ to the expected analytic solution.

Figure \ref{fig:code_val} panel $(a)$ shows the time evolution of $f(E_k,t)$ in the non-relativistic limit,
where all particles start with $\gamma \beta = 10^{-1}$ (or $E_k/m \simeq 0.005$) at random
directions. At $t=5$, which corresponds to a mean of scattering per particle $N_{\rm cpp}=3.3$, the 
derived $f(E_k,t)$ converges to the analytic solution, up to the expected random statistical noise
which becomes more significant at the lowest $E_k$ bins due to the small number of particles per bin. 
The bin sizes are of equal logarithmic size, to allow for the wide dynamic range in $E_k$ presented. 
To minimize the noise, we present the results only in the energy range where all consecutive bins have 
$N_{\rm bin}>10$.

Figure \ref{fig:code_val} panel $(b)$ shows the time evolution of $f(E_k,t)$ for a similar simulation 
in the relativistic regime, where the particles start with $\gamma\beta = 10$ ($E_k/m \simeq 9$).
As noted above (eq.~\ref{eq:rel_energy_gain}),  the energy exchange per scattering in the relativistic regime is much higher, so the relaxation of the DF is faster. In addition, 
already at $t=1$ the mean scattering per particle reaches $N_{\rm cpp}=5$, and therefore $F(E_k)$ relaxes 
to the expected analytic MJ solution. Note also the
break in the spectral index from the non-relativistic MB $F(E_k)\propto E_k^{1/2}$ 
at $E_k< 0.1$ to the relativistic MJ $F(E_k)\propto E_k^2$ at $E_k> 1$.

\subsubsection{\label{subsubsec:valid12}Non-identical particles}

Another validation test is for systems composed of two types of particles with an initial DF which is far from statistical equilibrium. 
Again we follow the time evolution of the DF, and test if it reaches the expected statistical equilibrium DF. 
The system consists of light $m_1=1$ and heavy $m_2=1836$ particles (similar to an electron + proton plasma).
The simulation initial condition is $\gamma\beta=10^{-3}$ for both types of particles. The system is far from
statistical equilibrium as the mean kinetic energy of the two particle populations is  $m_2/m_1=1836$, instead  of 
$1$ for equipartition. The number of light and heavy particles are $N=10^5$ each.
The colliding particles are randomly selected from the total list of $2\times 10^5$ particles, and the possible
collisions are therefore: heavy + heavy, heavy + light, and light + light particles. 

Figure \ref{fig:code_val_two_type} presents the time evolution of  $F(E_k)$ for the light and heavy particles. 
The DF start as a $\delta$ function, offset by a factor of $m_2/m_1=1836$ in energy. 
The high-mass particles lose energy, and the low-mass particle gain energy. By $t=10$, where $N_{\rm cpp}=0.03$, that is only 6\% of the particles were scattered, the shape of the two $F(E_k)$ of the scattered particles becomes quite close to MB, which are offset by the mass ratios.
The fast approach of each individual DF to a MB reflects the internal collisions of the equal
mass particles, which indeed establish equilibrium within each system following a few scattering. For example, at $t=550$ where $N_{\rm cpp}=3.7$ for the non-identical particles above, and at $t=5$ where $N_{\rm cpp}=3.3$
for the equal mass particles (Fig.~\ref{fig:code_val}).
The coupling of the two particle populations through collisions is much slower due to their large mass ratio. 
Since the energy gain per collision of the light particles is $\Delta E_{kl}\simeq 2m_l v_h^2$ (eq.~\ref{eq:non_rel_energy_gain}),
then the number of scattering required to reach equipartition with the heavy particles final energy (see below) 
of $\frac{1}{2}E_{kh}\simeq \frac{1}{4} m_h v_h^2$ is
$N_{\rm cpp}\simeq E_{kh}/\Delta E_{kl}=\frac{1}{8} m_h/m_l$, or $N_{\rm cpp}=1836/8=230$. Indeed, as Fig.~\ref{fig:code_val_two_type}
shows, at $t_{eq}=10000$, where $N_{\rm cpp}=480$, the two populations reach full equipartition, in contrast with $t=5$ 
where $N_{\rm cpp}=3.3$ is sufficient for full equipartition for the identical particles (Figure~\ref{fig:code_val}).

The total energy of the system, $E_t$, is
\begin{eqnarray}
E_t = N_1\times \langle E_1(m_1,T)\rangle + N_2\times \langle E_2(m_2,T)\rangle \ ,
\end{eqnarray}
where $N_1,N_2$ are the particle numbers of the two populations,  and $\langle E(m,T) \rangle$ is their mean energy
(eq.~\ref{eq:MeanEnergy}). Since $m_1\ll m_2$, and their initial velocity is the same, then initially $\langle E_1 \rangle\ll \langle E_2 \rangle $ and $E_t\simeq N_2\times \langle E_2 \rangle$.
In statistical equilibrium  $\langle  E_2\rangle\simeq \langle  E_1\rangle$, and since $N_1=N_2$ then 
$E_t\simeq 2N_2\times  \langle  E_2\rangle$, or a factor of two drop in the initial 
$T$ of the heavier particles, and a factor of $m_2/2m_1$ rise in the light particles initial $T$.

\begin{figure}[b]
\includegraphics[width=8 cm]{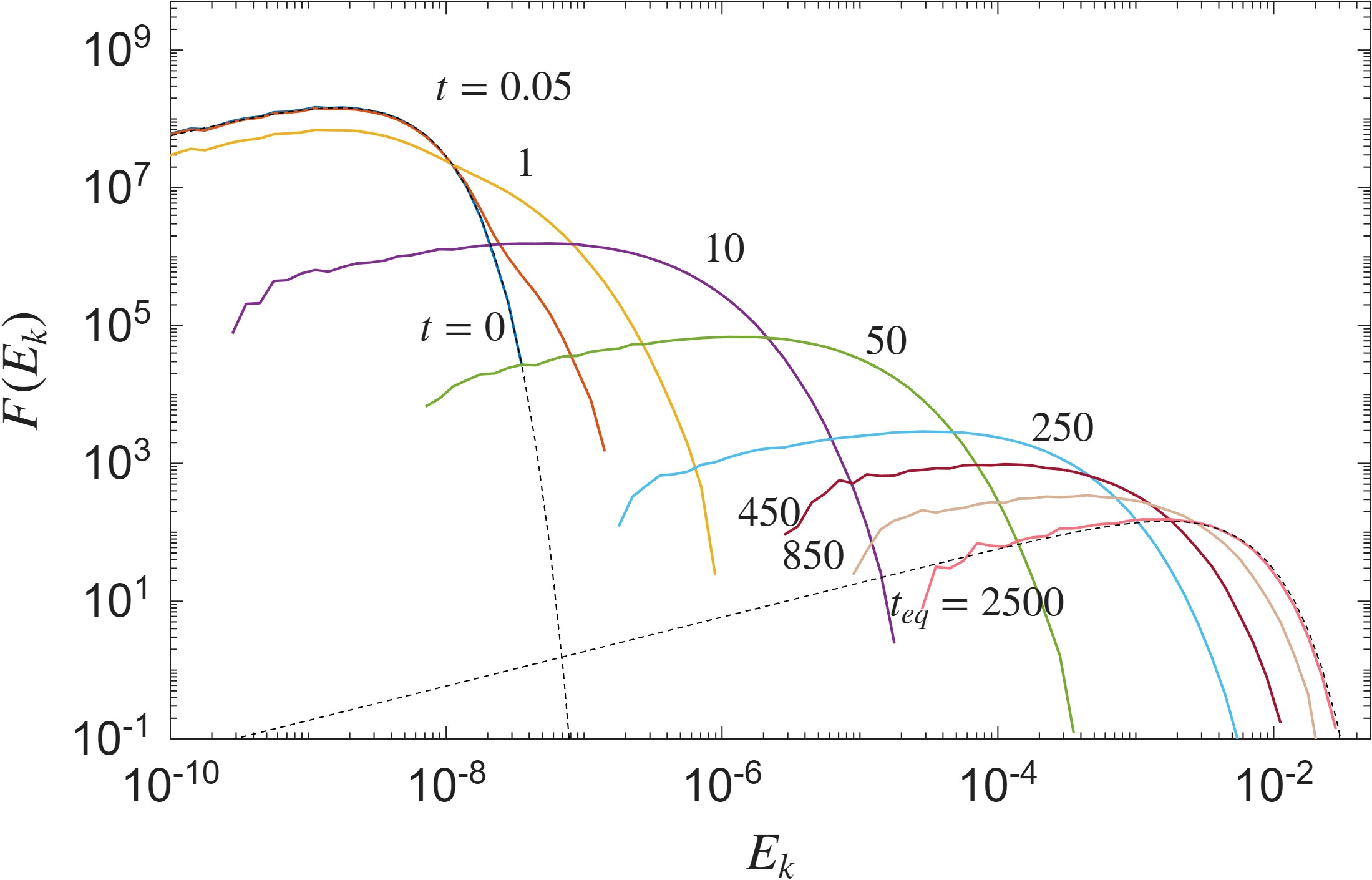}
\caption{\label{fig:whole_evol}The time evolution of low-mass particles DF as it approaches equilibrium with the background MB
of the high-mass particles. The energy distribution is depicted at various time steps, ranging from $t=0.05$ close to the initial MB DF (dashed line at $t=0$), until it reaches equilibrium at $t_{eq}=2500$ with MB DF of the background high-mass particles (dashed line).}
\end{figure}

The simulations above verify that the numerical algorithm reaches the expected equilibrium solution for equal and non equal mass particles, in the relativistic and non-relativistic limits. 

Note that the initial temperatures (i.e. mean energy) of the heavy and 
light particles set the energy exchange rate, and thus the equipartition timescale of the system. However, as demonstrated in the simulations below, the formation of the PL DF solution
is the intermediate asymptotic state solution, which occurs at the intermediate energies, that is when the system is far from the initial
state and from the final equipartition state. The nature of the PL DF (index and normalization), which is formed in steady state
at intermediate energies, is independent of the particles initial and final $T$ values.

\begin{figure*}[t]
\centering
\includegraphics[width=8 cm]{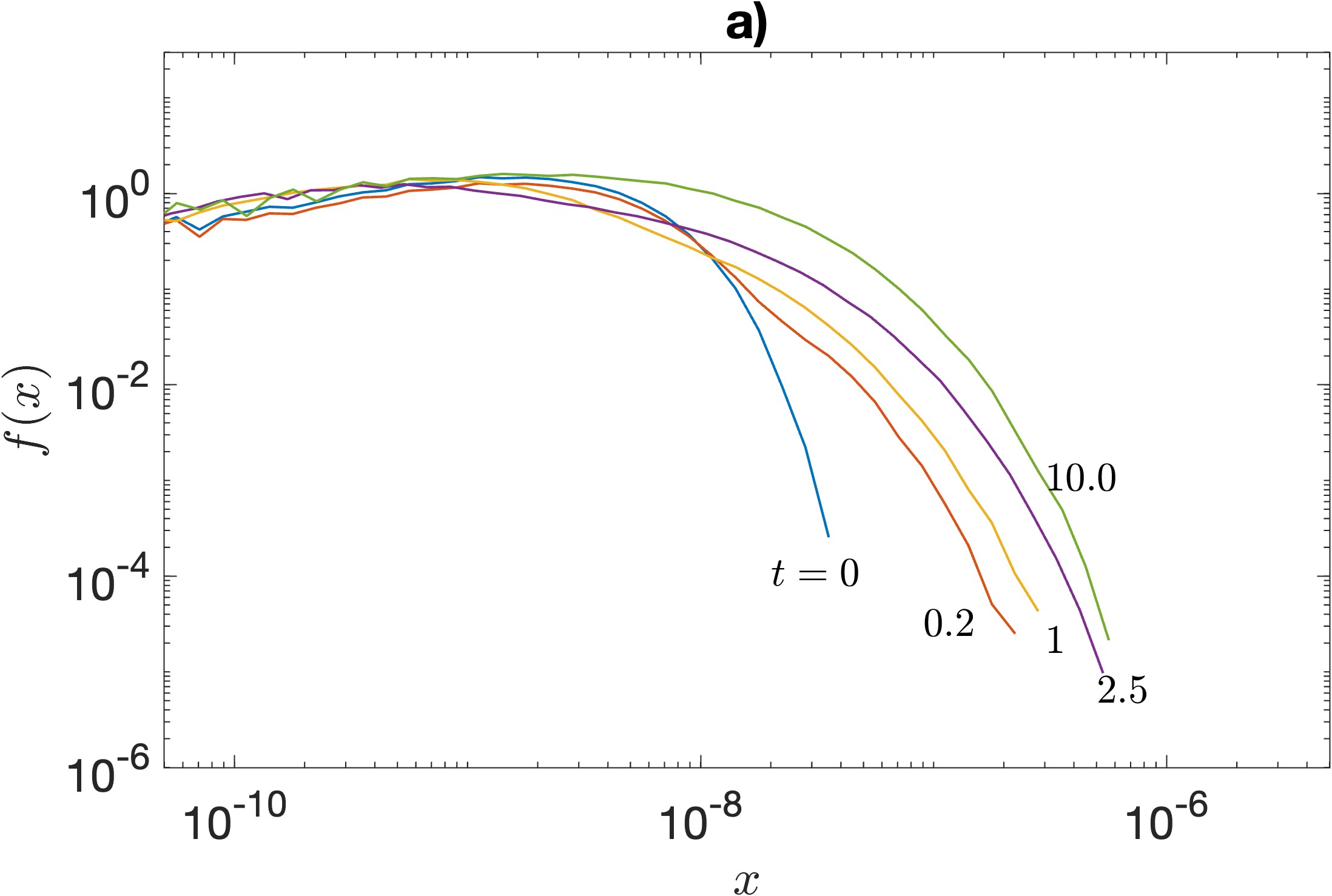}
\includegraphics[width=8 cm]{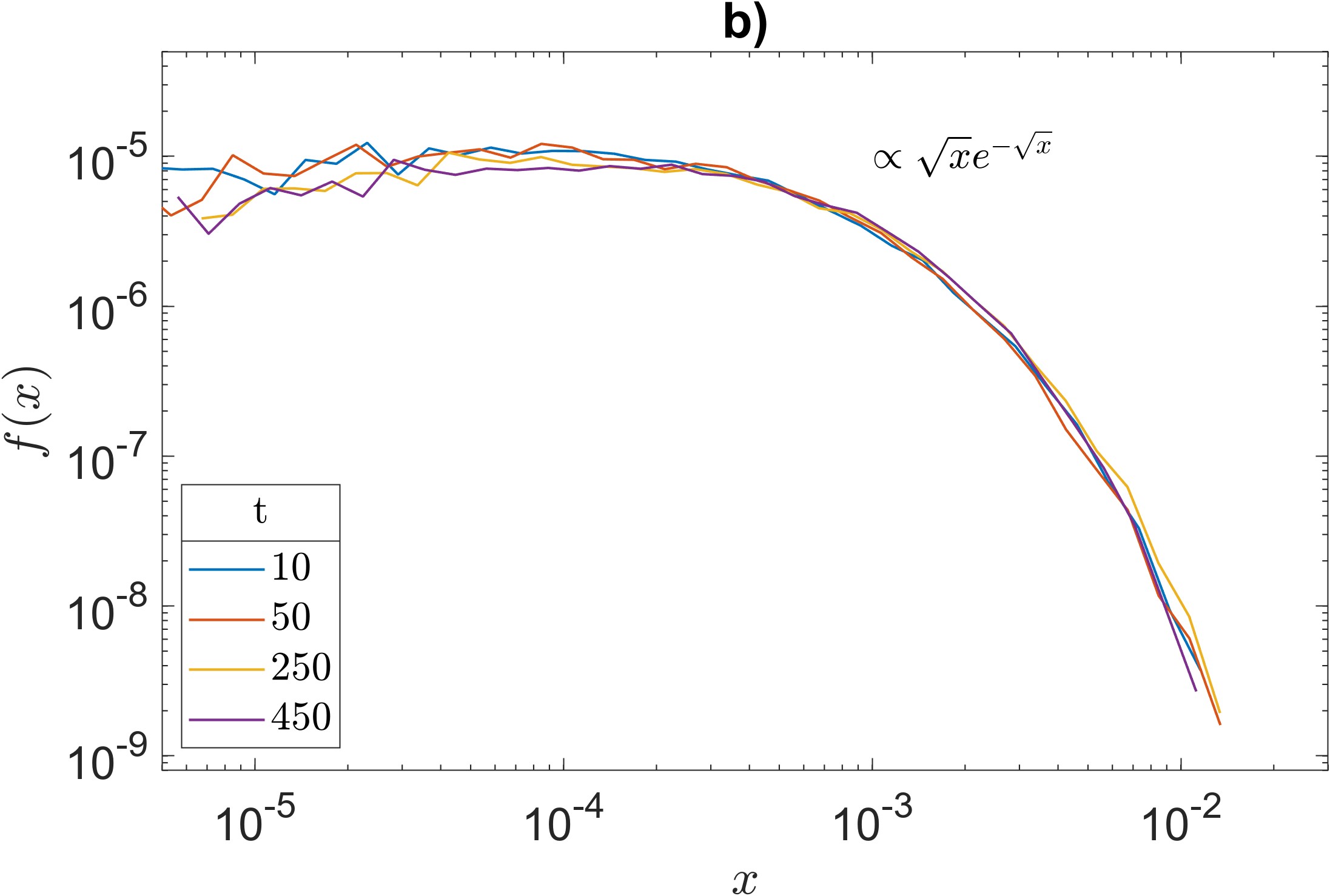}
\includegraphics[width=8 cm]{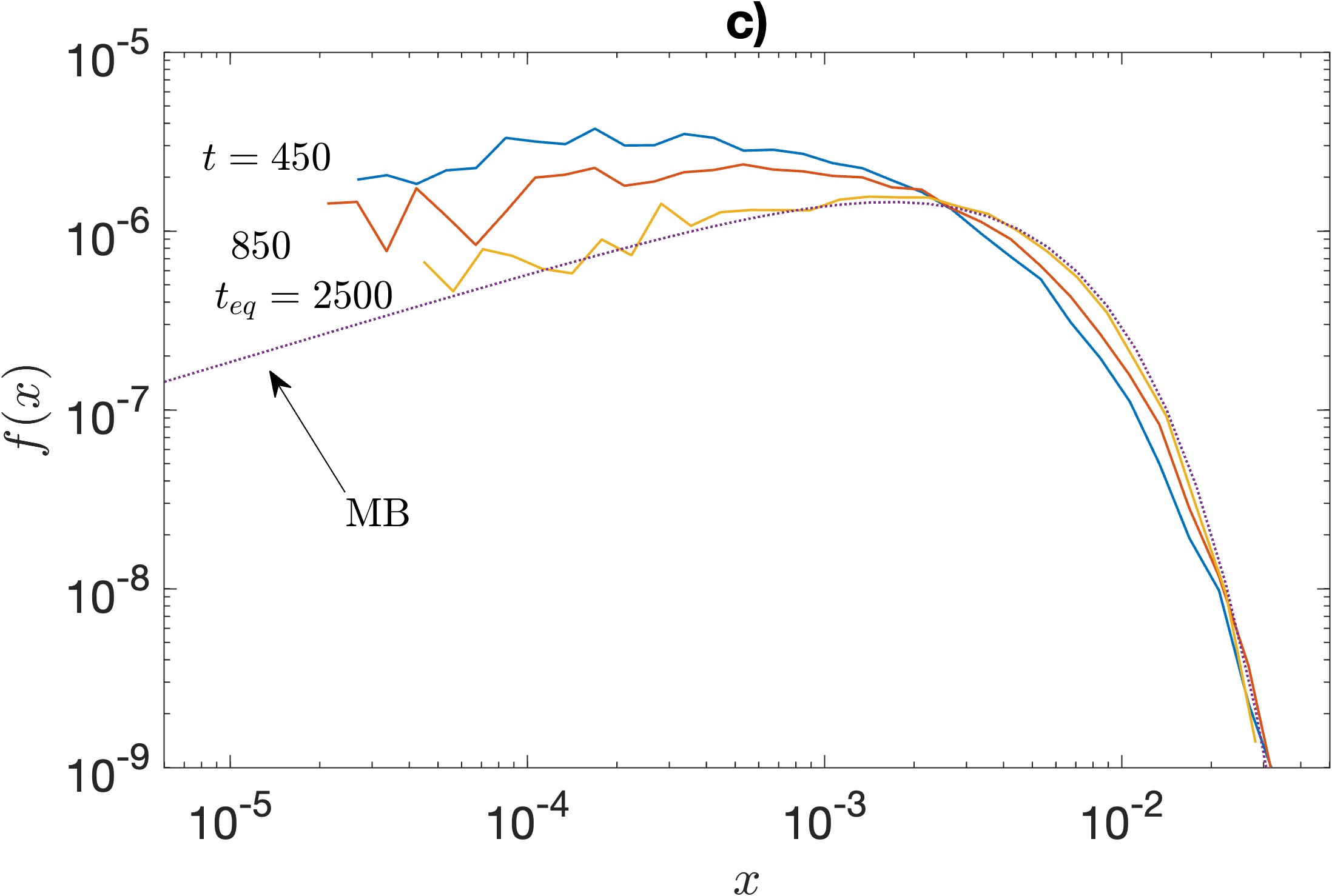}
\includegraphics[width=8 cm]{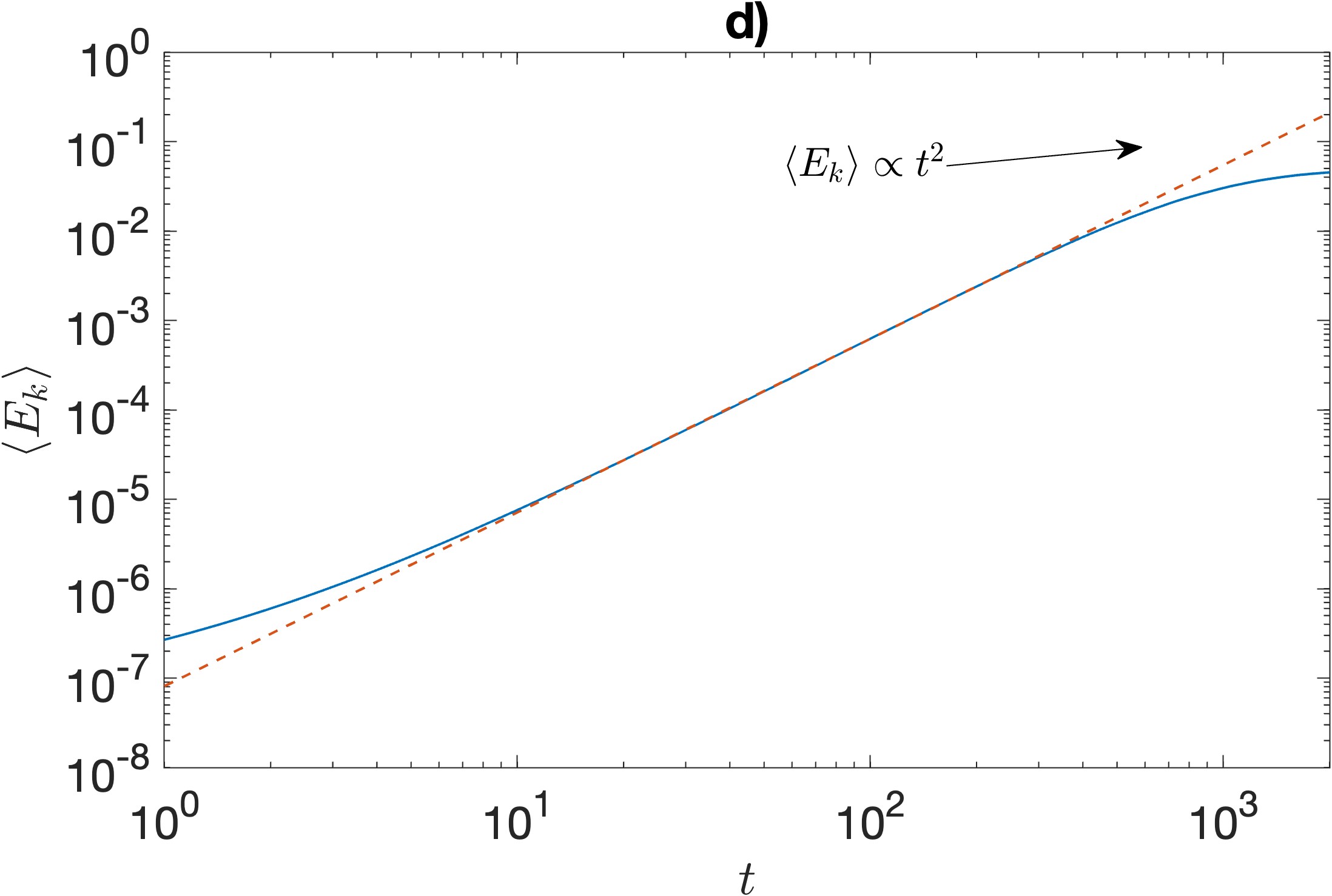}
\caption{\label{fig:steps_nop_time_dep} The formation of a self-similar DF asymptotically at the intermediate time steps.
The DF are each scaled by a factor $s$ according to $f(x)=f(s\cdot E)/s$, which satisfies $\int f(E)dE=1$. 
Panel \textbf{a)}, the DF shape evolution at $t=0-10$. 
The DF develops a more extended tail to higher energies, compared to the initial MB DF.
Panel \textbf{b)}, the evolution at the intermediate time interval, $t=10-450$. The DF reaches the intermediate asymptotic self-similar shape, $f(x)\propto \sqrt{x}e^{-\sqrt{x}}$. Panel \textbf{c)} the evolution at $t=450-2500$, where the DF of the low-mass particles evolves from the self-similar shape to a MB, where $f(x)\propto \sqrt{x}e^{-x}$.
 Panel \textbf{d)},  the time evolution of the mean kinetic energy $\langle E_k(t) \rangle$ (solid line). 
 The self-similar DF reaches asymptotically a dependence of $\langle E_k(t)\rangle\propto t^2$ (dashed line)
 at the intermediate time interval.}
\end{figure*}

\begin{figure}[b]
\includegraphics[width=8 cm]{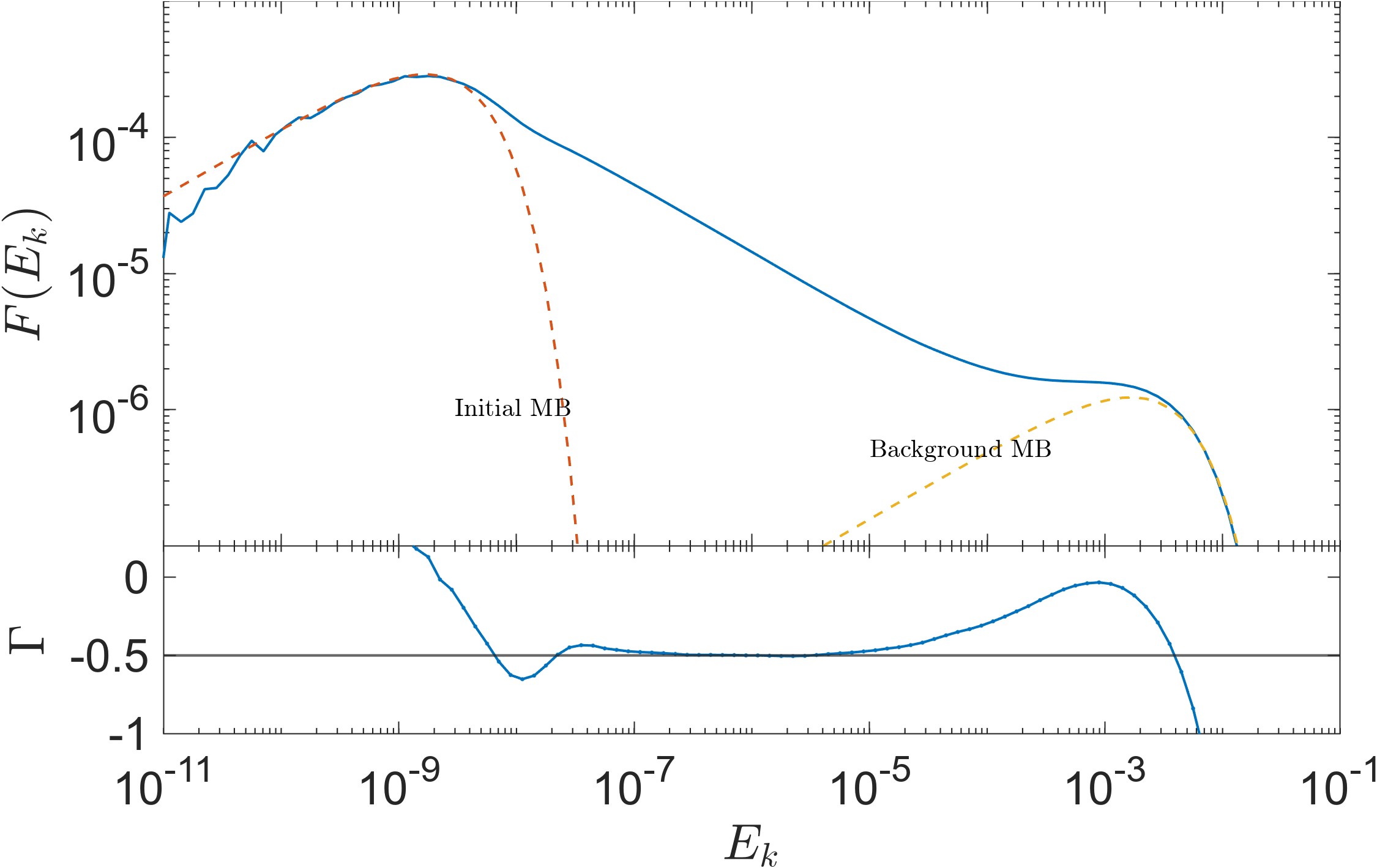}
\caption{\label{fig:stedy_state_injection} The steady-state DF $F(E_k)$, obtained for a constant injection rate of low energy particles. The DF follows a PL at $10^{-9}\ll E_k \ll 10^{-3}$, the intermediate energy range which is far from both the initial and final equilibrium energies. The DF PL index $\Gamma$ is displayed in the lower panel, which reaches asymptotically $\Gamma=-0.5$
in the intermediate regime. The scale-free PL solution is reached at intermediate energies far from
the initial and final energies, which are the characteristic quantities of the system.}
\end{figure}

\begin{figure*}
\includegraphics[width=8 cm]{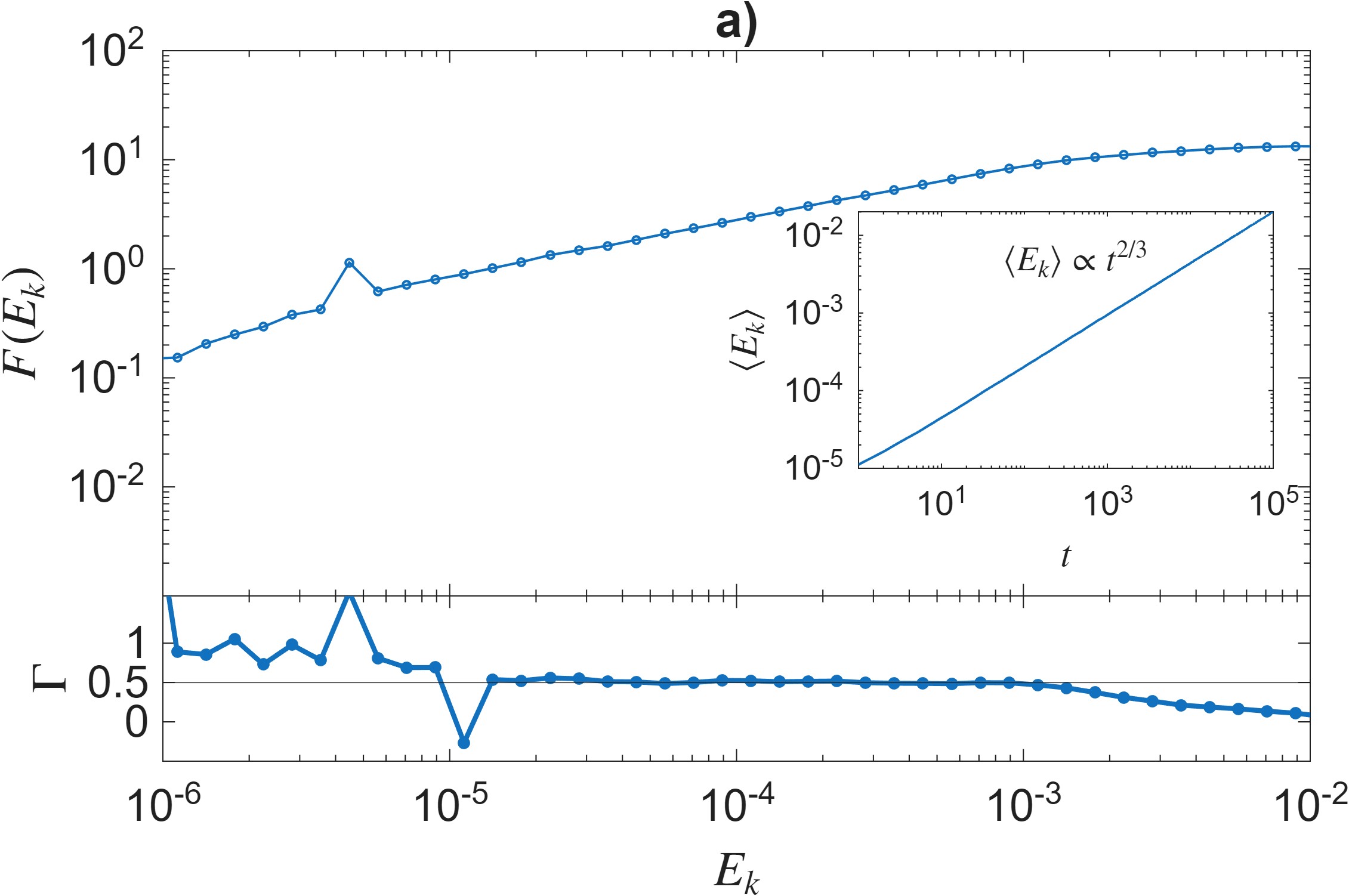}
\includegraphics[width=8 cm]{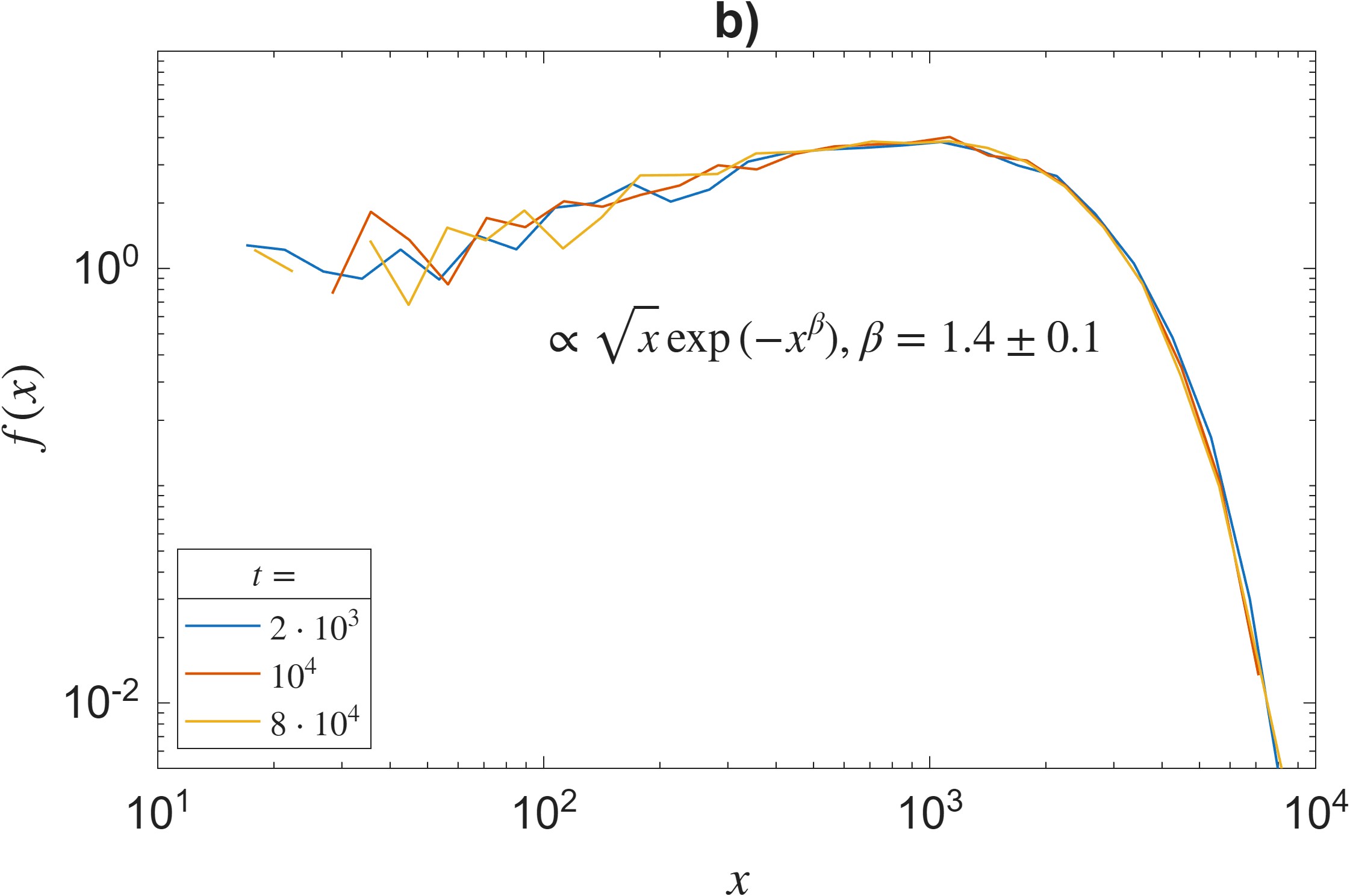}
\includegraphics[width=8 cm]{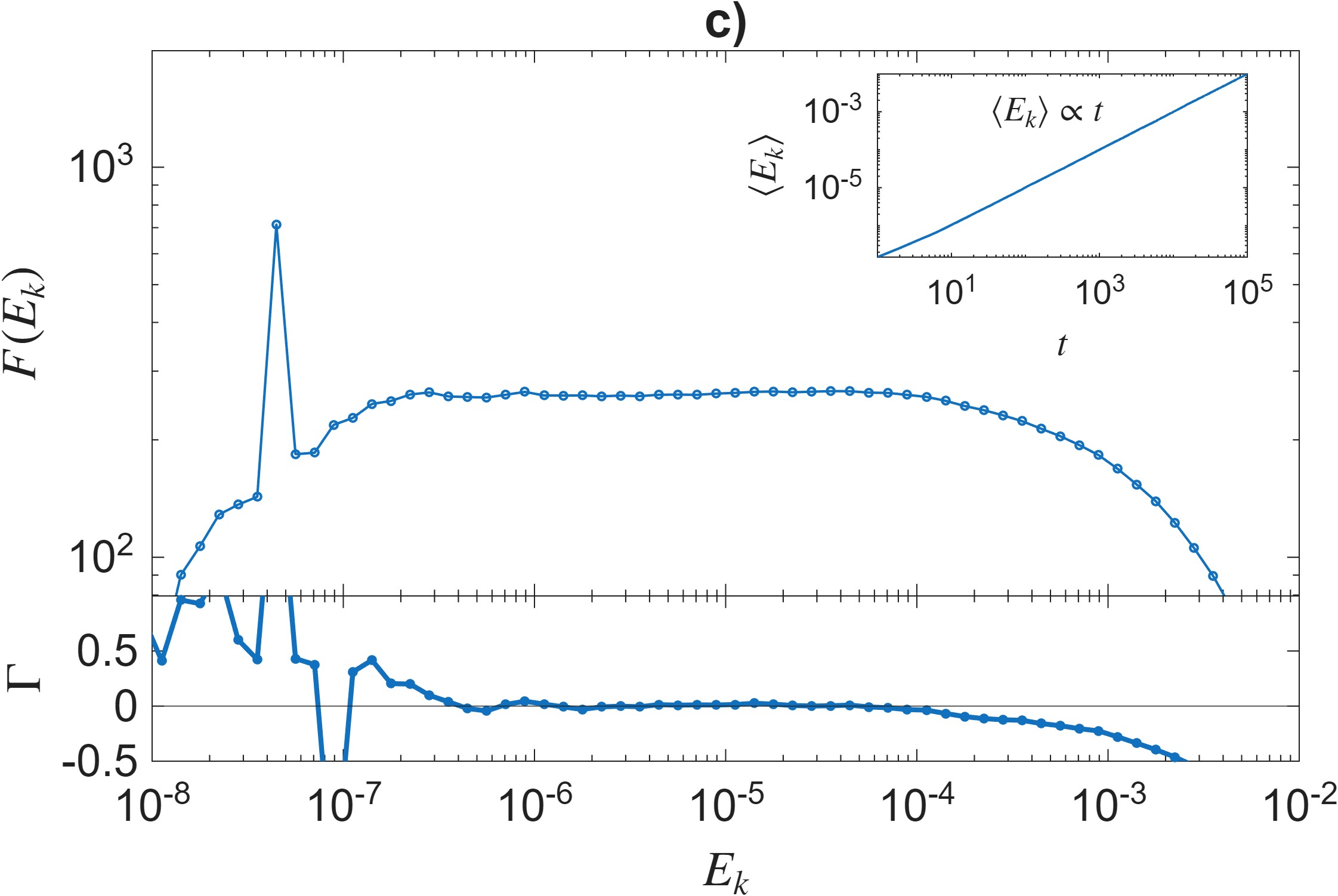}
\includegraphics[width=8 cm]{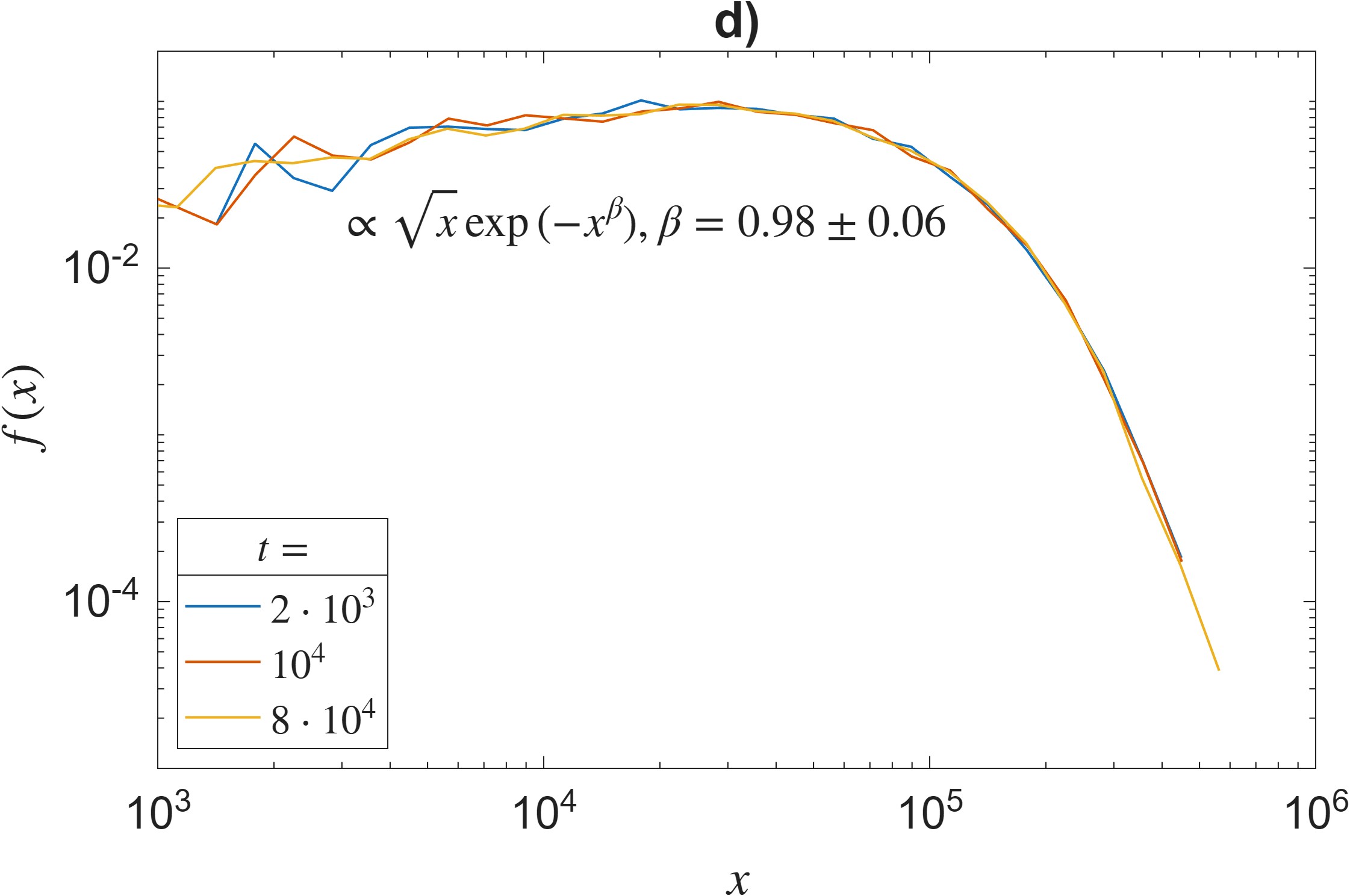}
\caption{\label{fig:clasic_evol_cross_section}  The steady-state DF for a constant injection rate, as in Figure~\ref{fig:stedy_state_injection}, but for two possible $\sigma(E_k)$.  \textbf{Panel} \textbf{a)}, for $\sigma(E_k) \propto E_k^{-1}$. The lower sub-panel displays the PL index, $\Gamma=0.5$ in the intermediate energy regime $10^{-5}<E_k<10^{-3}$. The embedded plot portrays the mean energy $\langle E_k(t) \rangle$, which follows $\langle E_k \rangle \propto t^{2/3}$.
\textbf{Panel} \textbf{b)}, the DF self-similarity at intermediate times where the PL solution holds. The best fit self-similar DF functional form
is given.
\textbf{Panels} \textbf{c)+d)} similar to panels \textbf{a)+b)} 
for $\sigma(E_k) \propto E_k^{-0.5}$. Here $\langle E_k \rangle\propto t$, $\Gamma=0$, and self-similarity holds with
a modified functional form.}
\end{figure*}

\section{\label{sec:results}Results}

Below, we explore the conditions which lead to the formation of a PL DF. We follow the time evolution of the DF in systems which are initially far from equilibrium. Each system is composed of two populations of particles. In the first case the particles have a mass ratio $m_2/m_1\gg 1$, with an initial $v_2/v_1 =1$. The collisions lead to 
a continuous acceleration of the light particles. In the second case $m_2/m_1= 1$, and initially $v_2/v_1\gg 1$. The collisions lead to an energy cascade of the fast particles, and to the formation of a shower of secondary particles. In the first case there is a continuous injection of light particles, and their energy grows either linearly (non-relativistic) or exponentially (relativistic), leading to a PL DF in energy. In the second case the number of particles grows exponentially, while their mean energy
drops exponentially, leading again to a PL DF for a constant injection of fast particles.

\subsection{\label{sec:asceleration}The acceleration algorithm}

Here we describe the simulation of the mix of light and massive particles. Since the energy loss 
of the massive particles is negligible, compared to the energy gain of the light particles
(section \ref{subsec:valid} and Figure ~\ref{fig:code_val_two_type}), we assume for simplicity 
that the DF of the massive particles remains constant. We can also neglect the light - light particle collisions,
which as shown below does not affect the final result. We therefore follow the evolution of the light
particles only, following their scattering with the massive particles.

The simulation includes $N=10^5$ massive particles, each with a mass $M=10^6$, which are non-relativistic  ($\gamma\beta=0.001$).
We let the massive particles collide with themselves a few times, 
to set an initial MB DF (which is kept constant), before they start colliding and accelerating the light particles. 
The simulation includes $N=10^5$ light particles with $m=1$, with the same initial $\gamma\beta=0.001$. The light particles also collide among themselves, to set up the initial MB DF.
This system is similar to the Lorentz gas of light particles embedded in a background of heavy particles \citep{Krapivskybook2010},
however the Lorentz gas model does not follow the light particles momentum evolution, and the eventual formation 
of a PL DF, as done here.

\subsubsection{Constant cross-section}
Here we follow the evolution of the light particles DF assuming a constant scattering cross-section, $\sigma(E) =\sigma_0$
(eq.~\ref{eq:R-freq}). 
Figure \ref{fig:whole_evol} presents the time evolution
of the light particles DF in kinetic energy space. The acceleration process exhibits three main time intervals. In the first interval, the MB DF transforms to a new DF. Once this transformation is completed, the intermediate time interval starts,
where the DF transforms in energy and amplitude, but keeps its shape in log-log scale, that is the DF remains self-similar. 
In the last interval, the DF relaxes back to the MB DF, but this time to the MB DF of the massive particles.
The light and heavy particles reach energy equipartition, that is the system reaches a full statistical equilibrium.

Figure \ref{fig:steps_nop_time_dep} provides a more detailed illustrations of the evolution of the DF in the initial, intermediate, and final time intervals. Panel 4.\textbf{a)} shows the DF at various times in the initial time interval, where each DF is shifted using a scaling transformation, $x = \alpha E$ and $y = \beta f(x)$. The parameters $\alpha$ and $\beta$ are derived at each time step to obtain the best overlap with the earlier timestep DF. The initial MB DF evolves and develops an increasing "tail" to high energies. The evolution of the DF is initially rapid, and gradually slows as the DF shape converges to a given form.
Panel 4.\textbf{b)} presents the DF shape at various times during the intermediate time interval. Clearly the DF approaches 
a self-similar form, with only minimal changes at the lowest and highest energies. The self-similar DF
is well fit by the simple relation
\begin{eqnarray}
    f(x) \propto \sqrt{x} \exp\left(-\sqrt{x}\right) ,
\end{eqnarray}
where $x=E/E_0(t)$. Panel 4.\textbf{c)}, presents the evolution of the DF in the last time interval, when the mean energy of the low-mass particles approaches the mean energy of the massive scatterers. The DF evolves in this time interval from the self-similar form back to a MB, $f(x) \propto \sqrt{x} \exp\left(-x\right)$,
this time to the one that matches the DF of the massive particles.
Panel 4.\textbf{d)} presents the evolution of the mean energy $\langle E_k\rangle$. The mean energy follows 
$\langle E_k\rangle\propto t^2$, as predicted by the analytical derivation for a constant cross-section $\sigma(E) = \sigma_0$ (eq.~\ref{eq:E_0t^2}).

Figure~\ref{fig:stedy_state_injection}, upper panel, presents the DF derived for a constant injection rate of $\dot{n}=1$ of light particles.
This is in contrast with Figure ~\ref{fig:whole_evol}, which presents the DF evolution for a single delta function
injection of light particles. Figure~\ref{fig:stedy_state_injection} is therefore effectively a sum of the DF
over a set of uniform time steps, from the current time backward by $\Delta t > t_{eq} = 2200$, which allows the first injected particles to reach equipartition with the massive particles. The DF forms a PL, with a low and a high energy turnovers. At $E_k<10^{-8}$ 
the DF is set by the initial MB of the injected particles, while at $E_k>10^{-3}$ it is set by the final equipartition 
MB of the background heavy particles. The amplitude of the final MB DF increases linearly with time, as the injected
particles accumulate there over time. The total integration time is set to $t > t_{eq}$ for presentation purpose to emphasize the accumulation effect and the formation of the final DF. Figure~\ref{fig:stedy_state_injection}, lower
panel, presents the local logarithmic slope of the DF $d \log F(E_k)/d\log E_k$. It demonstrates the "intermediate asymptotics" effect 
\citep[a term coined by][]{Barenblatt96} where the DF forms a PL with an index $\Gamma =-0.5$,  
which matches the index derived above analytically (eq.~\ref{eq:f0.5}). 

\subsubsection{Energy dependent collision cross-section}

Figure \ref{fig:clasic_evol_cross_section} explores the effect of an energy dependent 
cross-section, $\sigma (E)$, on the DF PL index in the intermediate energy regime, and also on the
shape of the self-similar DF at the intermediate time range. The simulation algorithm is similar
to the one described above for the constant cross-section case. Though here the initial DF of the 
total of $10^6$ light particles injected, is a delta function with $\gamma\beta=10^{-2}$ and randomly distributed
velocity directions, instead of a MB DF. Similarly, the background massive particles have a delta function velocity
distribution with $\gamma\beta=10^{-2}$ at random directions, instead of the MB DF. As shown below, 
the simulation results do not depend on the exact shape of the 
initial and background DFs.

Figure \ref{fig:clasic_evol_cross_section}.\textbf{a)} presents the simulation results for $\sigma (E) \propto E^{-1}$,
i.e. $\delta=-1$ (eq.~\ref{eq:sigmagamma}).
The upper panel shows that $F(E_k)$ forms a PL with $\Gamma=0.5$ (lower panel) in the intermediate energy range
$E_k=10^{-5}-10^{-3}$. As the inset shows, the mean kinetic energy scales as $\langle E_k \rangle \propto t^{2/3}$,
i.e. $b=2/3$, as expected from the analytic solution for $\delta=-1$ (eq.~\ref{eq:indexgamma}). 
The simulation PL index of $\Gamma=0.5$ matches the analytically derived index (eq.~\ref{eq:index.delta})
for $a=1$ (constant injection) and $\delta=-1$.

Figure \ref{fig:clasic_evol_cross_section}.\textbf{b)}, explores the self-similar part in the DF evolution
in the simulation above.
The DF indeed becomes self-similar on the intermediate energy timescales. The functional form of the DF is 
described well by the analytic expression $f(x) \propto \sqrt{x} \exp(-x^{1.4})$, in contrast with
$f(x) \propto \sqrt{x} \exp(-\sqrt{x})$ derived above for the $\delta=0$ case. We do not attempt here to derive $f(x)$ analytically. 
However, as noted above, the resulting DF PL index is independent of the specific functional form of the self-similar $f(x)$.

Figure \ref{fig:clasic_evol_cross_section}.\textbf{c)} and \ref{fig:clasic_evol_cross_section}.\textbf{d)},
show the simulation results for the $\sigma \propto E^{-0.5}$ case. This simulation leads to $\langle E_k \rangle \propto t$, and $\Gamma=0$, which again agree with the analytic derivations. Here the simulations give $f(x) \propto \sqrt{x} \exp(-x)$.

The three $\sigma$ cases explored above, $\sigma \propto E^{\delta}$ with $\delta=-1,-0.5,0$ lead to
three self-similar functions, $f(x) \propto \sqrt{x} \exp(-x^{\beta})$, with $\beta \simeq 1.5, 1, 0.5$.
suggesting the relation $\beta\simeq -\delta +1/2$. Thus, a steeper drop of $\sigma(E)$ leads to
a steeper exponential cutoff of $f(x)$.
Note that the amplitude of $\sigma$ is irrelevant to the solutions above, as it affects only the collision
rate (eq.~\ref{eq:f_c}), and thus the conversion from the dimensionless $t$ used here (which is a measure
of the number of collisions), to a physical time.  

\subsubsection{\label{sec:non_re_to_re_asc} The transition to the relativistic regime}

\begin{figure}
 \includegraphics[width=8 cm]{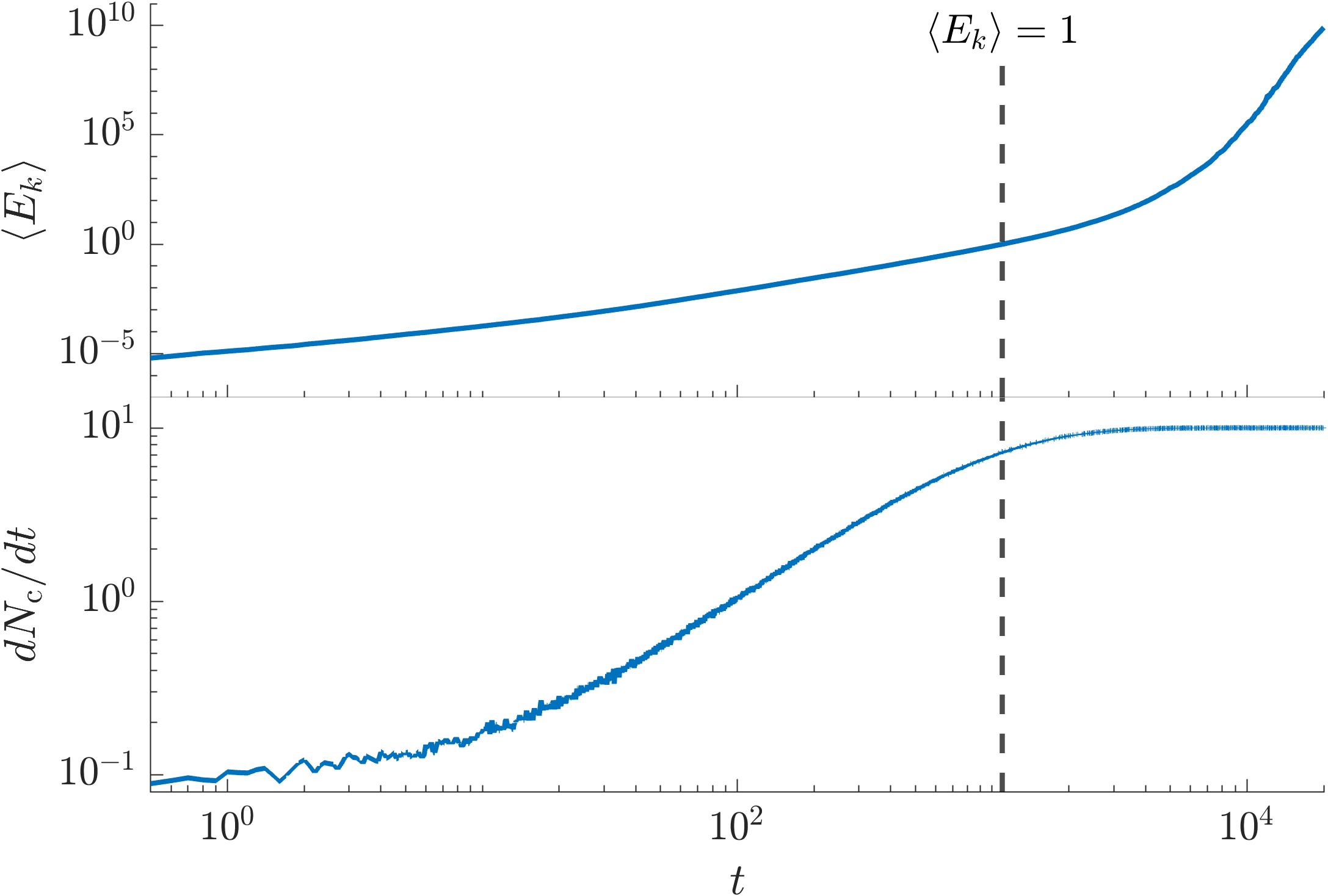}
\caption{\label{fig:E_evol}The time evolution of the mean kinetic energy $\langle E_{\rm k}(t)\rangle$ (upper panel), and the collision rate $dN_{\rm c}/dt$ (lower panel) for the simulation presented in Figure~\ref{fig:rel_evol}.
The vertical dashed line marks the transition time ($t=10^3$) from non-relativistic to relativistic, where $\langle E_{\rm k}\rangle = 1$.
At  $t< 10^3$ ($\langle E_{\rm k}\rangle < 1$), $\langle E_{\rm k}\rangle \propto t^2$, while at $t> 10^3$
($\langle E_{\rm k}\rangle > 1$) $\langle E_{\rm k}\rangle \propto e^t$.
The collision rate follows $dN_{\rm c}/dt \propto t$ in the asymptotic non-relativistic regime as the velocity grows linearly with time, and it approaches a constant value for the relativistic regime, where the velocity approaches $c$.
}
\end{figure}

\begin{figure}
\includegraphics[width=8 cm]{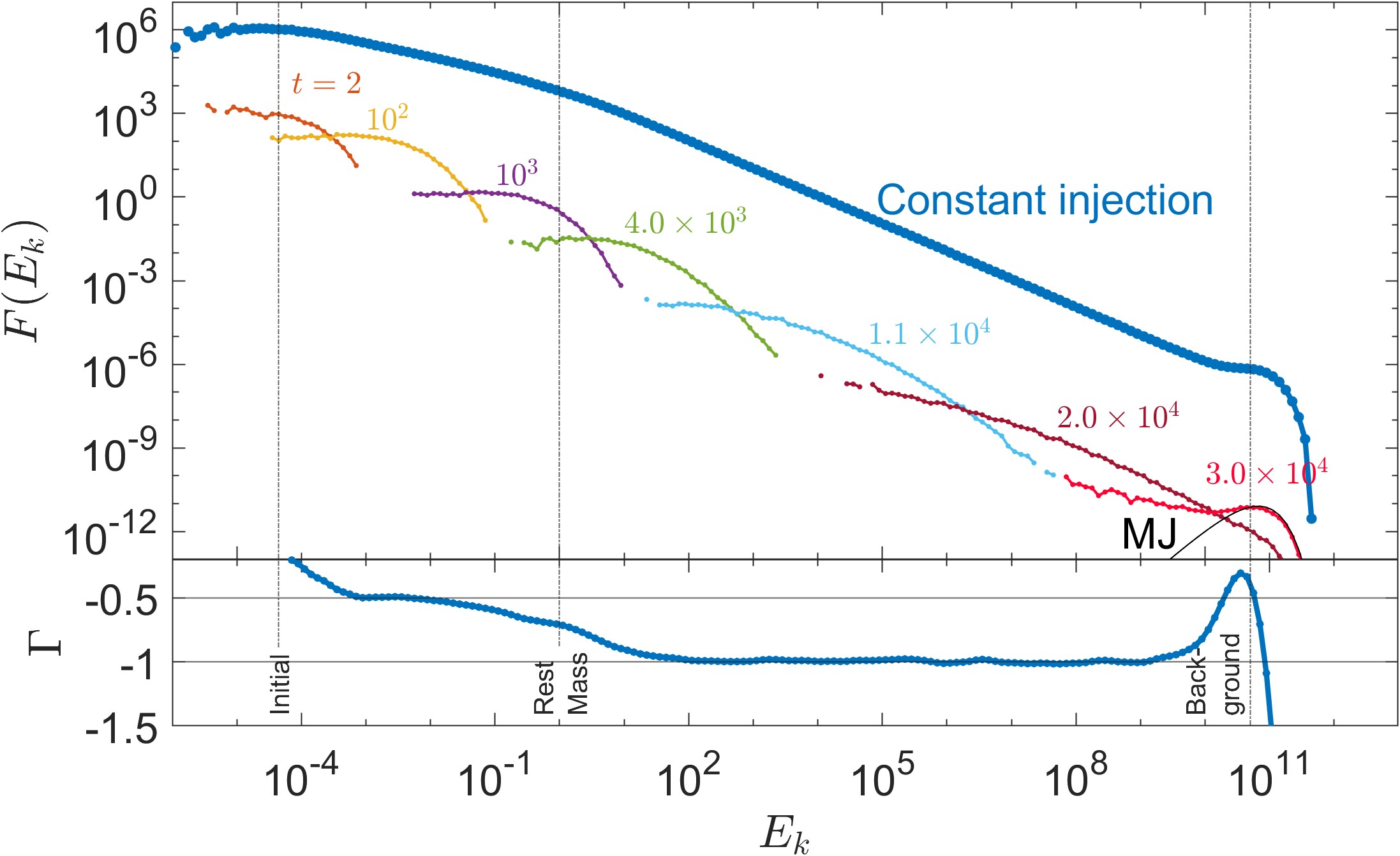}
\caption{\label{fig:rel_evol} The time evolution of the DF for particles injected in the non-relativistic regime
($E_k=10^{-4}$), and reaches equilibrium with the heavy particles in the ultra-relativistic regime ($E_k=10^{10}$). The DF for a steady state constant injection is displayed in the blue line. The lower panel presents  $\Gamma$ in the steady state solution, 
which switches from $\Gamma=-0.5$ at $E_k\ll 1$ to $\Gamma=-1$ at $E_k\gg 1$. Since $E_k=1$ is a characteristic energy scale, 
where the dynamics switches from non-relativistic to relativistic, the scale-free  PL solutions apply only
at the two separate regimes, $10^{-4}\ll E_k\ll 1$ and at $1\ll E_k\ll 10^{10}$.}
\end{figure}

The simulations above assume non-relativistic dynamics, that is $E_k\ll m$ for both the light and heavy particles.
Here, we follow the evolution of the low-mass particles DF from the non-relativistic to the ultra-relativistic regime. 
The heavy particles mass is $m_h=10^{15}$ with an initial momentum per unit mass
of $p/m=\gamma\beta=10^{-2}$, and therefore $E_{kh}\sim 5\times 10^{10}$. The heavy particles collide among themselves and reach
a MB DF, which remains unchanged (that is an effectively infinite background of heavy particles). The light particles have $m_l=1$, and the same initial momentum per unit mass, $\gamma\beta=10^{-2}$ or $E_{kl}=5\times10^{-5}$, and are all injected into the system with
this energy, moving at random directions. Upon repeated scattering $E_{kl}\rightarrow E_{kh}$, and since $E_{kh}\gg m_l$ the light particles eventually become ultra-relativistic as they evolve towards equipartition with the heavy particles. In contrast, the heavy particles remain non-relativistic. The light particles undergo scattering with the heavy particles only (assumed to be an infinite pool), with a constant $\sigma$. 

Figure \ref{fig:E_evol}, upper panel, presents the time evolution of the light particles mean kinetic energy $\langle E_{\rm k} \rangle$.
At $t<10^3$ $\langle E_{\rm k} \rangle<1$, the particles are non-relativistic and $\langle E_{\rm k} \rangle\propto t^2$ (see Figure \ref{fig:steps_nop_time_dep}),
while at $t>10^3$ $\langle E_{\rm k} \rangle>1$ we get the expected exponential growth, which starts flattening at $t>10^4$ as $\langle E_{\rm k} \rangle$
starts approaching equipartition with the heavy particles. The lower panel presents the collision rate $dN_{\rm c}/dt$,
which approaches $dN_{\rm c}/dt\propto t$. This occurs since $dN_{\rm c}/dt\propto v$, and $\langle E_{\rm k} \rangle\propto v^2 \propto t^2$, or
$v \propto t$ when $\langle E_{\rm k} \rangle\gg E_{\rm inj}$. At $t>10^3$
$v \rightarrow c$ and the collision rate approaches a constant. Clearly, $\langle E_{\rm k} \rangle=1$ sets a characteristic energy scale,
and self-similar solutions are expected at $\langle E_{\rm k} \rangle\ll 1 $ and $\langle E_{\rm k} \rangle\gg 1 $.

Figure \ref{fig:rel_evol}, upper panel, presents the time evolution of the light particles DF following an injection event at $t=0$ up to $t=3\times 10^4$ (colored lines). The figure also shows the DF for a constant injection rate $\dot{n}=1$
(blue dotted line). The lower panel, presents the steady state $\Gamma$ as a function of $E_{kl}$.
The DF becomes a PL with $\Gamma=-0.5$ at $E_{kl}\sim  10^{-3}- 10^{-1}$, as expected in the non-relativistic regime.
At $E_{kl}\sim  10^{-1}- 10^{1}$ $\Gamma$ steepens gradually from $-0.5$ to $-1$, while in the ultra-relativistic regime 
$E_{kl}\sim  10^{1} - 10^{9}$ we get $\Gamma=-1$, before the light particles reach equipartition with the heavy
particles at $E_{kl}\sim  5\times 10^{10}$. Clearly, $E_{kl}=m_l$ sets a specific energy scale at $E_{kl}=1$ in the system dynamics. The intermediate asymptotic PL solutions hold only at energies which are well away from the three
specific scales, $E_{kl}=5\times 10^{-5}, 1$ and $5\times 10^{10}$. The DF transforms from $F(E_k)\propto E_k^{-1/2}$
in the non-relativistic regime $5\times 10^{-5}\ll E_{k}\ll 1$, to $F(E_k)\propto E_k^{-1}$ in the ultra-relativistic regime
$1 \ll E_{k}\ll 5\times 10^{10}$.

We note that the $\Gamma=-1$ derived from the ultra-relativistic simulation above agrees with the analytic 
mixed state solution derived above (eq.~\ref{eq:mixed-state}). This solution applies here as the number of
particles grows linearly with time (constant injection rate, $a=1$), while their mean energy grows exponentially (eq.~\ref{eq:rel_energy_gain}).

\subsubsection{\label{sec:Similarity_relativistic}The self-similarity transformation in the relativistic regime}

In the relativistic regime the system dynamics remains scale-free, but the self-similar form of the distribution function differs qualitatively from the non-relativistic case. The energy growth is exponential, rather than polynomial in time, leading to an exponential self-similarity in the time evolving DF. The self-similar form of the DF, $f(x)$, is not derived by a multiplicative
scaling of the energy, $x=E/E_0$, but rather by an exponential scaling $x=exp(E/E_0)$. This exponential 
transformation is motivated by the exponential growth of $E_0(t)$ in the relativistic regime (eq.~\ref{eq:exponential}).

\begin{figure}[b]
\includegraphics[width=8 cm]{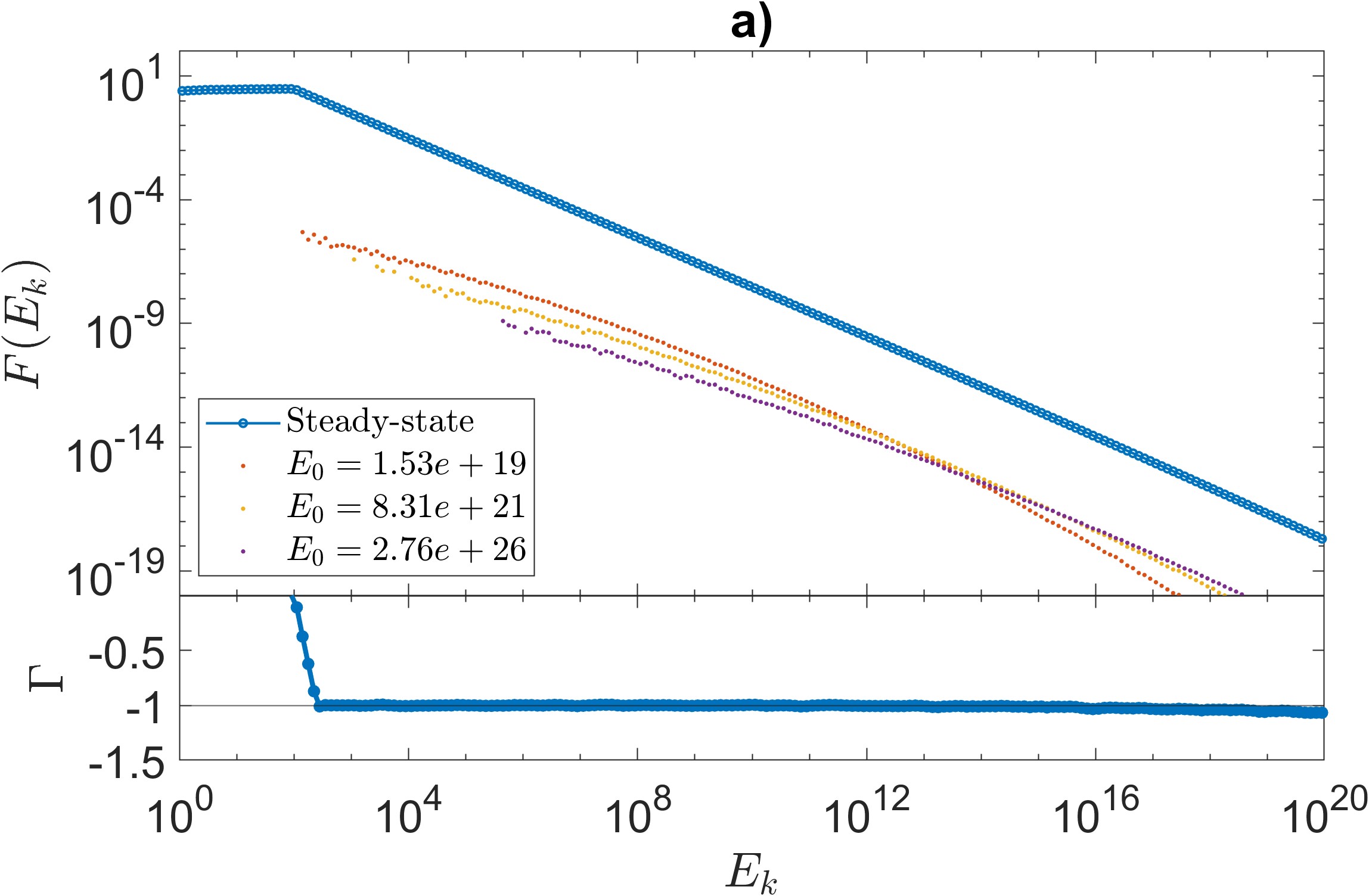}
\includegraphics[width=8 cm]{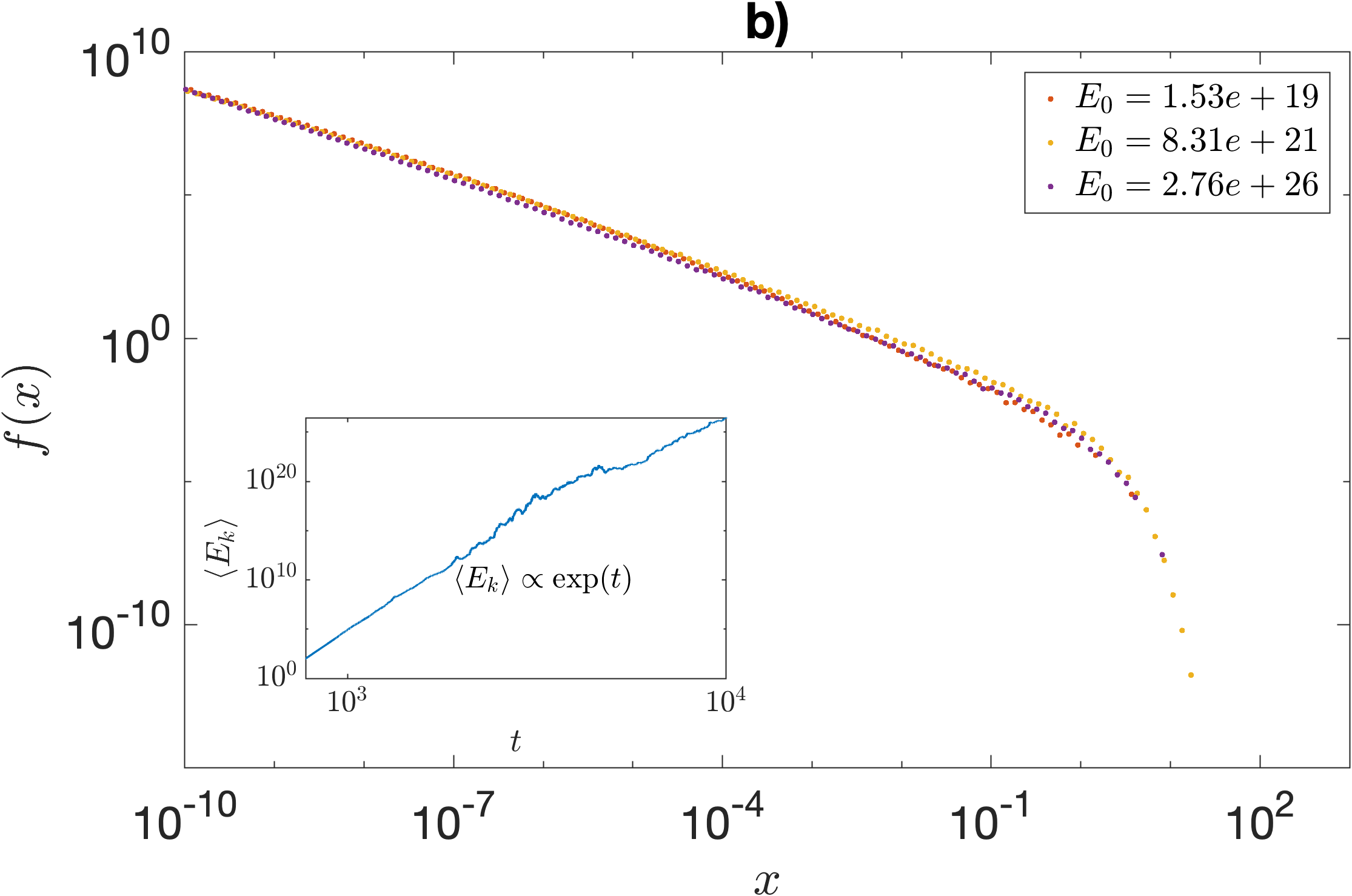}
\caption{\label{fig:relativistic_evolution_and_self_similarities} 
The steady-state DF for a constant injection rate with a constant cross-section in the ultra-relativistic regime. 
\textbf{Panel (a)}: Three snapshots of the DF following a single injection event at three different times.
The blue curve shows the DF resulting from a constant injection rate. The lower sub-panel presents the corresponding $\Gamma$, illustrating the formation of the expected $\Gamma=-1$ PL for exponential self-similarity and a constant particle injection rate. 
\textbf{Panel (b)}: The exponential self-similarity of the three DF from panel (a), transformed according to Eq.~\ref{eq:transformation}. Note the extremely wide dynamic range in $x$ where self-similarity holds.}

\end{figure}

Figure~\ref{fig:relativistic_evolution_and_self_similarities} presents a more detailed look at the 
exponential self-similarity. It shows a simulation similar to the one presented above, with a 
background of heavy particles with $\gamma\beta=10^{-2}$, but this time with an infinite mass, to avoid
a characteristic maximal energy. The light particles are injected here 
with $\gamma \beta = 100$, that is well into the relativistic regime . 

Figure~\ref{fig:relativistic_evolution_and_self_similarities}, upper panel, shows the time evolution of the energy DF for three snapshots which span a mean energy increase by seven orders of magnitude, along with the steady-state DF obtained for a constant injection rate. The DF exhibits a well-defined PL tail with an index $\Gamma = -1$, consistent with the analytical expectation derived in Section~\ref{sec:analytic} for a mixed exponential system, where $N_p(t) \propto t$, and $E_0(t) \propto \exp(\beta t)$. The inset displays the exponential growth of the characteristic energy with time.

Figure~\ref{fig:relativistic_evolution_and_self_similarities}, lower panel, presents 
the self-similar, $f(x)$, derived from the three DF presented in the upper panels.
The exponential self-similarity is derived from the following transformation:
\begin{eqnarray}
\label{eq:transformation}
    x = \exp\left(\frac{E}{E_0}\right), \quad
    f'(x) = \frac{f(E) E_0}{\exp\left(\frac{E}{E_0}\right)},
\end{eqnarray}

These results demonstrate that exponential self-similarity naturally arises in relativistic systems with scale-free dynamics and steady-state injection, and leads to robust PL solutions across a wide energy range.

\subsection{\label{sec:Cascade}The Cascade (shower) algorithm}

Below, we simulate the DF evolution in another system of colliding hard spheres which is far from statistical  equilibrium.
Above we simulated a mix of light and heavy particles with the same initial velocity, and thus different energies.
Here we simulate a mix of initially fast and slow particles with the same mass, and thus different energies.
The energetic particles cascade down in energy,
producing a shower of secondary fast particles, which further cascade down in energy, until the system reaches equipartition.
As above, we follow the DF evolution when the two populations are non-relativistic, and 
for a mix of non-relativistic and ultra-relativistic particles.

In contrast with the acceleration simulation above, where the number of the initially slow injected particles is conserved,
and their energy increase, here the number of fast particles grows as the shower develops, but the total particles energy is conserved (in the self-similar regime, far from equipartition). At each collision a background slow particle becomes a fast particle, which doubles the number of fast particles within a collision time. As a result, the number of particles grows exponentially,
while their mean energy falls exponentially. Given the scale-free scattering dynamics, i.e. lack 
of a characteristic energy when the scattering is either relativistic or non-relativistic, 
a self-similar time evolving DF is formed, leading to a PL DF for a steady state injection rate. 
Self-similarity breaks when the fast particles mean energy approaches the background particles mean energy.

The cascade simulation below demonstrates the formation of a PL DF in an exponentially  evolving system where both 
$N_p(t)\propto e^{\alpha t}$ and $E_0(t)\propto e^{\beta t}$ in the relativistic regime (see section~\ref{sec:analytic} above), in contrast with
the acceleration simulations above, where $N_p(t)\propto t^a$ and $E_0(t)\propto t^b$ in the non-relativistic regime, or the
mixed state $N_p(t)\propto t$ and $E_0(t)\propto e^{\beta t}$ which holds in the relativistic regime, also leading to a PL DF. 

The low velocity particles are assumed to originate from an infinite pool of thermal background particles. 
As a result, scattering among high-energy particles are assumed to be negligible, compared to their scattering
with the thermal background. We also do not follow scattering among the background 
low-energy particles, since these scattering just maintain their MB distributions, 
which does not change given the infinite size of this background population. We follow only
the scattering between high-energy and low-energy background particles. The high-energy particles include both the initially injected
high-energy particles, and the low-energy thermal particles following their collision with a high-energy particle.
As a result, the high-energy particle population grows exponentially and inevitably becomes too large to handle
numerically within a few scattering times. To sustain the simulation, we introduce a population reduction procedure. 
In this procedure we follow the true size of the population as a function of time, but take a randomly selected subsample
each time the population reaches some size limit, say we randomly select $10^4$ particles each time the simulated population reaches $10^5$. 
This procedure follows the time evolution of the DF, 
to an accuracy set by the population size which is being followed.

\subsubsection{\label{sec:Cascade_non_relativistic} Particle cascade in the non-relativistic regime}

In the non-relativistic simulation we inject a single high-energy particle with $E_k=5\times10^{-1}$, which collides with
a randomly selected low-energy particle drawn from a MB DF with a $\langle E_k\rangle=5\times10^{-9}$.
The collision is realized based on the collision rate, i.e. the collision probability, as described above (section ~\ref{chap:outlines}, eq.~\ref{eq:R-freq}).  We use a constant cross-section $\sigma_0$. Following a collision, the low-energy particle is incorporated into the array of high-energy particles, which are followed in the simulation. The duration of the time step is calculated based on the maximal collision rate, as in the acceleration simulations. 

\begin{figure}[b]
\includegraphics[width=8 cm]{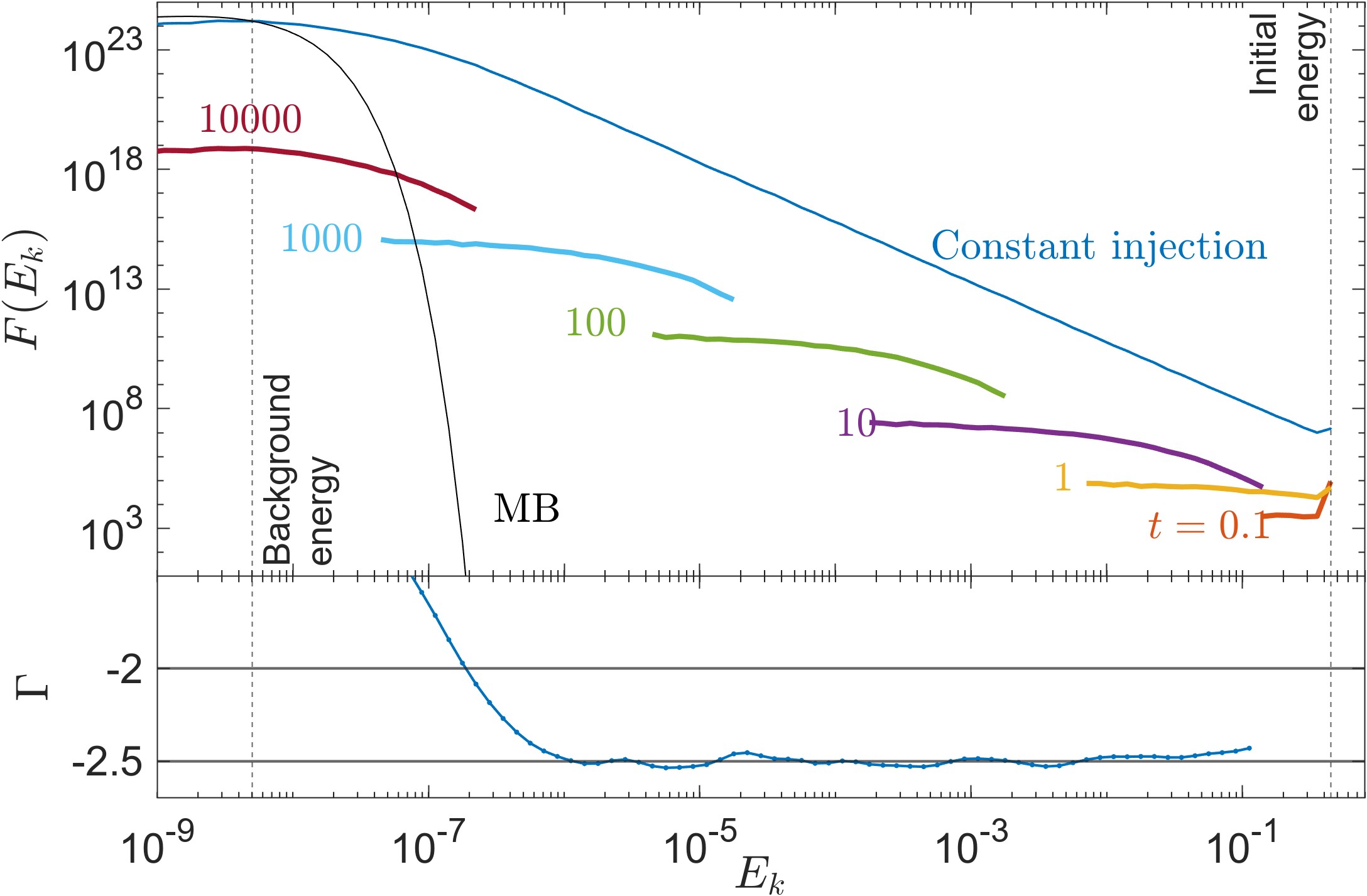}
\includegraphics[width=8 cm]{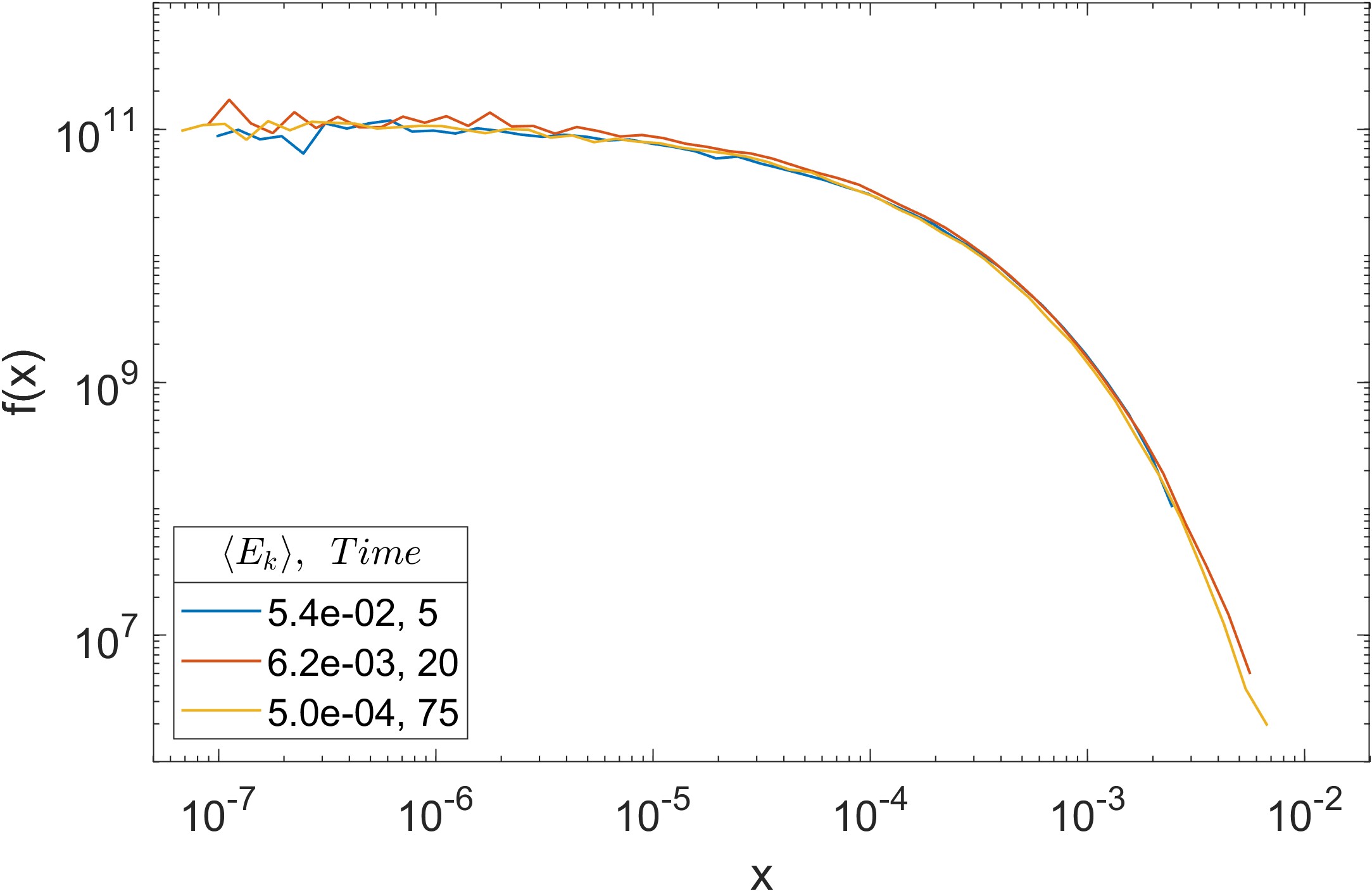}
\caption{\label{fig:cascade_clasic_evol} The non-relativistic cascade simulation. \textbf{Upper panel} The time evolution of the DF of particles injected with $E_k=0.5$ in a thermal low energy background with a $\langle E_k\rangle =5\times 10^{-8}$. The solid blue line shows the steady state DF for a constant injection, which leads to the formation of a $\Gamma= -2.5$ PL at the intermediate energy regime $5\times 10^{-8}\ll E_k\ll 0.5$ (lower sub-panel). \textbf{Lower panel} The DF self-similarity in the intermediate energy regime
at three different times. This simulation demonstrates the formation of a PL DF when the system dynamics leads to
$N_p(t)\propto e^{\alpha t}$ and $E_0(t)\propto e^{\beta t}$.}
\end{figure}

Figure \ref{fig:cascade_clasic_evol}, upper panel, presents the high-energy particle DF evolution with time following a single injection event, from $t=0.1$ to $10^4$. The number of particles increases
with time, and their mean energy drops. We follow the evolution until the particle mean energy is less than ten times the background mean energy.
Following that the particles reach thermalization with the background low-energy thermal population.

Figure \ref{fig:cascade_clasic_evol} also presents the DF derived for a constant injection rate, derived 
by summing up the DF of single injections at different times.  Since the high-energy particles produced during the shower do not interact with each other, the steady state solution can be derived by a superposition of an evolving DF following a single injection.
The steady state DF forms a PL with a constant $\Gamma = -2.5$, which starts to deviate due to the tail of the MB DF already
at $E_k<10^{-6}$, or a factor of $\sim 100$ larger than the thermal $E_k\sim 10^{-8}$. This transition is set by the specific number
of high-energy particles which accumulate at the thermal reservoir, a number which grows linearly with time for a constant injection rate.

Figure \ref{fig:cascade_clasic_evol}, lower panel, demonstrates the self-similarity of the time evolving DF produced by the cascade following a single injection at different times.
The self-similarity holds at intermediate times where the DF resides well away from the characteristic kinetic
energies, that is at $5\times10^{-9}\ll E_k\ll 5\times10^{-1}$. Since the total energy of the particles is conserved (equals the energy
of the injected particle), as long as the energy of the background particles is negligible, the self-similarity satisfies 
$(E_k,F(E_k))\rightarrow (sE_k, F(E_k)/s^2)$, in contrast with $(E_k,F(E_k))\rightarrow (sE_k, F(E_k)/s)$ when the number
of particles is conserved (in the acceleration scheme above).

\subsubsection{\label{chap:Cascade_relativistic} Particle cascade in the ultra-relativistic regime.}

Below we simulate the cascade process when the injected particle is in the ultra-relativistic regime, that is $E_k \gg m$.
The relativistic collision rate differs from the non-relativistic case (eq.~\ref{eq:R-freq}), which affects the DF 
PL index. We employ an algorithm similar to the one used in the previous section. 
Both the high-energy particles and the background particles have the same rest mass of $m=1$. The particles are injected 
at random directions with $\gamma \beta=10^7$. The background MB thermal particles have $\langle E_k\rangle =5\times10^{-9}$,
as in the non-relativistic simulation.

\begin{figure}
\includegraphics[width=8 cm]{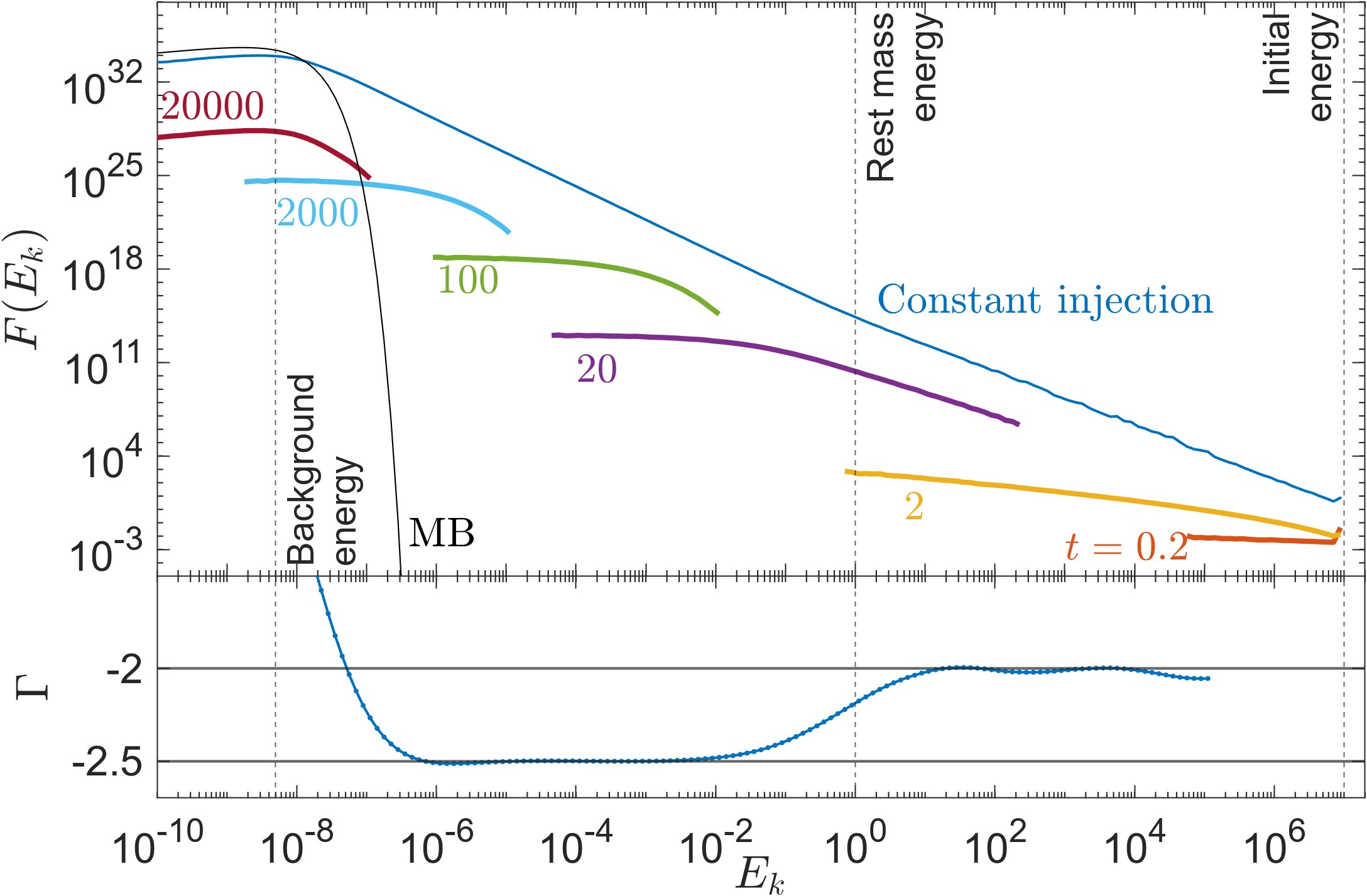}
\includegraphics[width=8 cm]{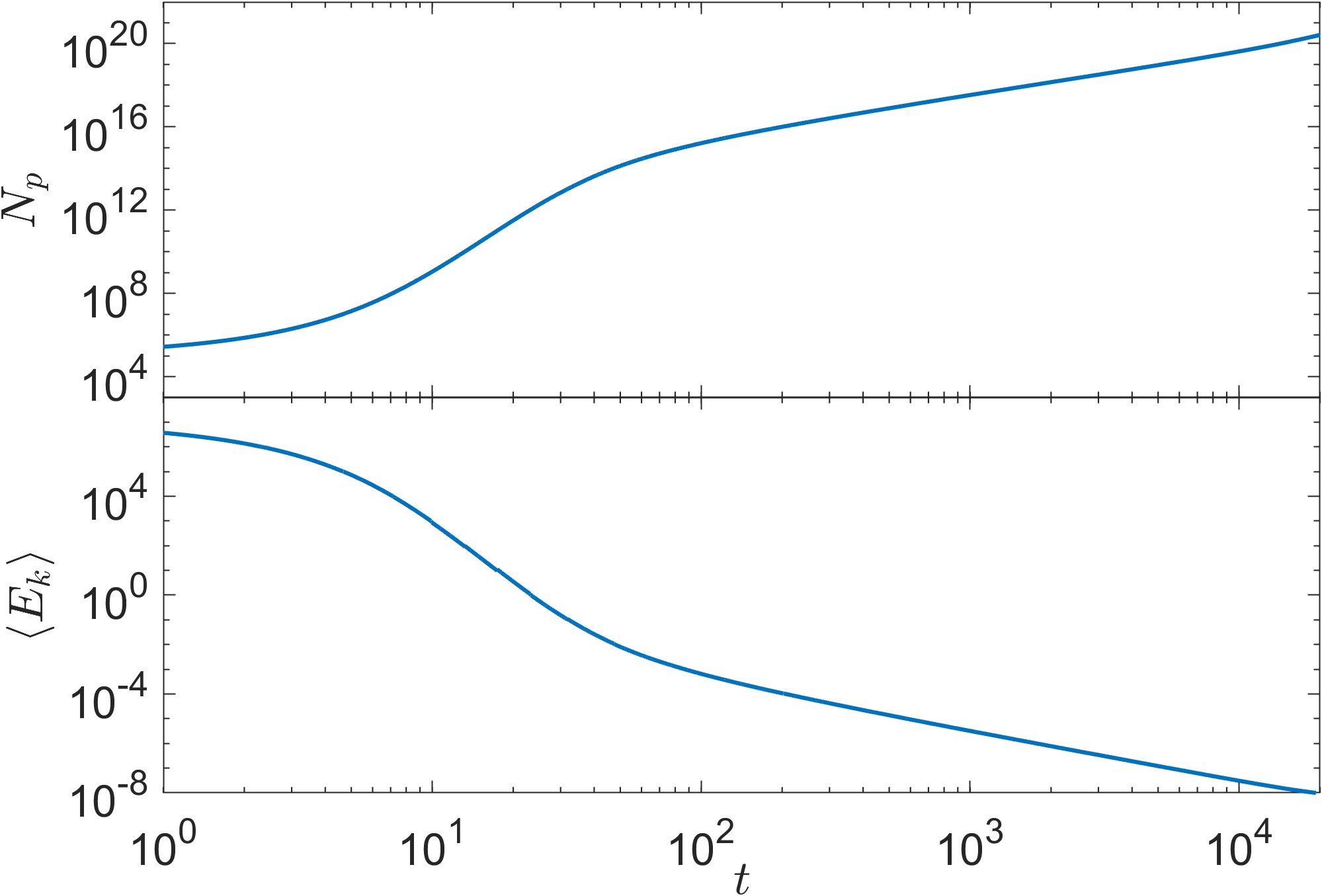}
\caption{\label{fig:cascade_evol_rel} \textbf{Upper panel} As in Figure ~\ref{fig:cascade_clasic_evol}, but for the injection of ultra-relativistic particles, which cascade down to the non-relativistic regime.
The constant injection now leads to a broken PL DF (lower panel) with $\Gamma=-2$ at $E_k>10$ and $\Gamma=-2.5$ at $E_k<0.1$. 
As in Figure~\ref{fig:rel_evol}, the scattering change from non-relativistic ($E_k\propto p^2$) to relativistic ($E_k\propto p$)
at $E_k=1$, which leads to the spectral index break at this energy. \textbf{Lower panel}: The increase in $N_p(t)$
and decrease in $\langle E_k(t)\rangle$ with time. The growth in $N_p(t)$ is exponential in the ultra-relativistic regime, and switches to a PL growth in the non-relativistic regime. Since the total kinetic energy is conserved in the cascade
process (as long as the background particles energy is negligible), we get $\langle E_k\rangle \propto N_p^{-1}$. }
\end{figure}

Figure \ref{fig:cascade_evol_rel}, upper panel,
presents the DF evolution from $t=0.2$ to $2\times 10^4$ (time calculated according to 
eq.~\ref{eq:time_incriment}). It is noteworthy that already at $t=2$ the tail of the DF extends down to the non-relativistic regime, indicating the rapid evolution of the DF in the relativistic regime compared to the non-relativistic region. This is also reflected by the much larger drop in
$\langle E_k\rangle$ by about a factor of $\sim 10^{16}$ in the ultra-relativistic simulation, compared to a drop of $\sim 10^8$
in the non-relativistic simulation, for comparable evolution times. The steady state DF, derived as above by integrating the individual
DF at equal time steps, shows the expected index break at $E_k\sim 1$, transitioning from the ultra-relativistic $\Gamma = -2$ to the non-relativistic index
of $\Gamma = -2.5$. The $\Gamma = -2$ index in the relativistic regime is simple to understand. Since $dN_c/dt$ is constant (for constant $\sigma$, see above following eq.~\ref{eq:E_0t^2}), then $N_p(t)\propto e^{\alpha t}$ and $\langle E_{\rm k}(t) \rangle\propto e^{\beta t}$ (note that $\langle E_{\rm k}(t) \rangle$ scales as the characteristic energy $E_0(t)$), 
and since $N_p(t)\times \langle E_{\rm k}(t) \rangle$ is constant (the total energy in the system), then $\alpha + \beta=0$, or $\alpha/\beta=-1$ which gives $\Gamma = -2$ (eq.~\ref{eq:index.greek}). 
When the cascade becomes non-relativistic $dN_c/dt\propto t$, leading to a more elaborate solution for 
$\Gamma$.

Figure \ref{fig:cascade_evol_rel}, lower panel, displays the time evolution of the particle number $N_p(t)$ and their mean energy $\langle E_k (t)\rangle$ ($\propto 1/N_p(t)$). 
The initial growth rate of $N_p(t)$ is exponential, and transforms into a PL growth at $t>30$ when the typical particle energy becomes non-relativistic.  
The plot presents the absolute particle number accumulated in the full cascade process, which reaches $\sim 10^{20}$. This number results from the energy of a single particle injected, which is $E_k\simeq 10^7$, with $\sim 10^4$ 
particles injected by the time $t\sim 10^4$, which gives a total injected kinetic energy $E_k\simeq 10^{11}$.
This energy is spread over $\sim 10^{20}$ particles, with an energy per particle which drops to $E_k\simeq 10^{-9}$,
which equals the mean background particles thermal energy, and the PL DF merges to the background MB DF.
Clearly, no numerical simulation can follow $N_p\sim 10^{20}$,
and the numerical reduction scheme, described above, handles this challenge and allows us to follow accurately the 
time evolution of the DF, regardless of the number of particles involved in the shower. 

\section{Discussion}

We can now answer the question posed in the title of the paper, \textit{when does a Maxwell-Boltzmann DF transforms into a Power-Law DF?}
The simulations above indicate there are three sufficient conditions: 1. The system starts far from equilibrium, e.g. low and high mass particles
with similar velocity, or low and high energy particles with similar mass.
2. The system dynamics is scale-free, that is there is no characteristic velocity/energy in the system.
The results of scattering events depend only on the relative velocity, and not on their 
absolute values. The scale-free  dynamics holds in the intermediate asymptotic regime away from the characteristic system scales, 
such as the initial energy, the final equilibrium energy, and the rest mass energy. 
The scale-free  dynamics leads to the formation of a time evolving self-similar DF.
3. The system is open with a scale-free  boundary condition, such as a steady state injection of particles
which are far from equilibrium into the system. The superposition of the self-similar DF of particles injected over time 
leads to the formation of a PL DF. In contrast, a closed system eventually reaches statistical equilibrium 
on a relaxation time, regardless of the initial conditions.

The specific form of the scale-free function $f(x)$ is set by the system dynamics, such as the 
gain or loss of the mean energy, relativistic or non-relativistic energies, the energy dependence of the scattering rate, etc'.
However, this specific form is integrated out when the steady state DF is derived, and the PL index
$\Gamma$ is set by the time dependence of the fundamental properties of the system, such as the particle
injection rate, and by the time dependence of the characteristic property of $f(x)$, specifically
$E_0(t)$ in the simulations above, which is set by the system dynamics. 

It may therefore be relatively straightforward to derive the expected DF PL $\Gamma$ in steady state 
for a given physical system, if conditions 1-3 above hold, without a need for a direct numerical simulation.

The system of colliding hard spheres is simplistic, and is selected exactly for this reason.
Below we briefly describe a number of more elaborate physical systems in nature which show a PL DF, and discuss 
whether the conditions described above are relevant there.

\subsection{Power-Law Formation by Shocks.} 

Relativistic electrons with a PL DF in energy are commonly observed in shocks, produced by e.g. supernova
or jets, as indicated by the synchrotron emission from the shocked medium. The origin of the PL DF is the
elastic scattering of charged particles by the ambient magnetic field irregularities, either between the shocked
and unshocked media (Fermi first order acceleration), or irregularities within the shocked gas (Fermi second order,
or diffusive shock acceleration). The formation of the PL DF is derived either with the particle in cell
simulations, which follow individual particles and their collective effects, or using magnetohydrodynamics
simulations which follows the gas global properties together with the Fokker--Planck diffusion approximation which follows
the particle DF evolution \cite{Drury1983, Jones1994, Achterberg2001, Lemoine2006, Lemoine2025}. 

The essence of the shock acceleration is the particle scattering by the magnetic field irregularities.
This scattering is effectively similar to the light particle scattering by the heavy particle 
in our simulations above, which leads to a PL DF. In astrophysical systems there are additional factors,
such as the particles escape fraction following a scattering event, energy loss between scattering events,
etc', but as long as these effects are scale-free, the DF solution in steady state is expected to remains a PL.

\subsection{Power-Law Formation by Energy Cascade}

In the particle simulations above we followed the formation of a PL DF in systems where high-energy particles are continuously injected into a colder medium and lose energy through collisions and the subsequent formation of a shower of secondary particles. A similar process occurs in solar flares, where 
high-energy electrons are injected into the chromosphere and are slowed by Coulomb collisions, forming a PL electron spectrum that produces the observed X-ray PL emission \citep{Brown1971,Bastian1998}. In cosmic-ray transport, continuous injection from sources such as supernova remnants, combined with synchrotron, inverse-Compton, and Coulomb losses, leads to a stationary electron spectrum that is a power law, with the index determined by the dominant loss mechanism \citep{Sarazin1999,Cowley2009}. \citet{ParkPetrosian1995} provide PL analytic solutions for Fokker–Planck equations, confirming that steady injection and scale-free transport yield stationary PL distributions over the intermediate energy range. Laboratory experiments
 demonstrate that continuous injection of high-speed neutral atoms into a cold plasma with inelastic collisional losses produce a stationary PL velocity distribution \citep[e.g.][]{Aranson2002},
consistent with predictions of energy-loss cascade models \citep[e.g.][]{Ben2005, Fujii2023}.

\subsection{Comptonization in hot Plasma}

PL photon spectra, $n(\nu) \propto \nu^{-\alpha}$, naturally emerge from Comptonization of soft photons in hot plasmas, as repeated scattering produce scale-free energy diffusion. The PL is produced in the regime of unsaturated Comptonization, that is when the photons do not scatter enough times to reach equipartition with the hot plasma. There is  
continuous injection of soft-photons into the hot plasma, and escape of high-energy photons with a constant escape 
probability, set by the system optical depth \citep{SunyaevTitarchuk1980, RL79}. The situation is analogous 
to the low energy light particles and high energy heavy particle scattering simulation above, though here the
light particles are massless, and therefore relativistic. The escape of the light particles in this simulation 
is scale-free, as the electron scattering opacity is independent of energy, and therefore the PL solution
still holds, but its index depends on the escape fraction \citep[eq.7.45b there]{RL79}

\subsection{Other systems with PL DFs}

As noted above, PL DF are prevalent in nature and characterize 
a great variety of systems \citep[e.g.][]{newman05, Stumpf2012critical}. Below we briefly discuss some mechanisms proposed,
and their relation to the general conditions proposed here. We stress again that we do 
not refer to PL relations derived from dimensional analysis, for example, that a pendulum osscilation  
time scales as its (length)$^{1/2}$. Or PL relations derived from the geometry of the problem, for example that the distance traveled
in a random walk scales like (number of steps)$^{1/2}$. We also do not refer to the PL relation between
two physical quantities, such as the allometric scalings in biological systems, e.g. 
that the metabolic rates of organisms scale as their (mass)$^{3/4}$ \citep{West1997allometric}.

A ubiquitous PL DF is the Pareto power-law of wealth distribution \citep{gabaix09}, which can be modeled 
by the Kinetic Exchange Models (KEMs) in
econophysics. This model uses the kinetic energy exchange of colliding molecules to simulate the derived
kinetic energy DF, which is assumed to be analogous to the dynamics leading to the observed wealth DF. 
These simulations show that a PL is formed by
a superposition of heterogenous KEMs \citep{Patriarca13}, a process which is formed in our simulations by the superposition of the time 
evolving self-similar DFs.

Another phenomena associated with a PL DF is a world wide pandemic, where the number of countries 
with a given number of cases follows a PL DF.
This can be understood by the exponential growth with time of the number of infected countries,
together with the exponential growth with time in the number of cases per country, as recently  demonstrated for the Covid pandemic \citep{Blasius20}. This process is similar to cascade simulations above, where both the energy per particle, and number 
of particles, evolves exponentially with time, which leads to a PL in the distribution of particles
(countries) per energy (Covid cases per country).

PL distributions are also commonly found in network systems, which show a PL DF in the number of connections per node.
The network can be for example a social network on the internet, computer nodes on the internet, or film actor
collaborations in movies. A commonly assumed mechanism is preferential attachment, whereby new nodes are more 
likely to connect to existing nodes with higher degree of nodes. The underlying dynamics, introduced in \citet{barabasi1999emergence}, is $dk_i/dt\propto k_i^\alpha$ where $k_i(t)$ is the degree of node $i$,
and $\alpha$ defines the attachment nonlinearity. This scale-free dynamics leads to a PL DF with $P(k) \sim k^{-\gamma}$.
We note in passing that a PL DF is also observed in the mass distribution of a large variety of systems, ranging in size from nanometer to mega light years.
We discuss in a follow up paper a possible mechanism which leads to the rather universal index in these systems.

\section{Conclusions}

Above we solved through direct Monte Carlo simulations some dynamics which leads to the formation of a PL DF in a system
of colliding hard sphere. The simple dynamics allows to solve the time evolution of the energy DF fully and accurately. We found there are three sufficient conditions: 1. The system is far from statistical equilibrium. 2. The dynamics is scale-free, and 3. The system is open with a steady state (or generally scale-free ) boundary condition. 

Conditions 1 and 2 with a delta function boundary condition (say a single injection event) 
lead to the formation of a time dependent DF with a self-similar form $f(x)$, where $x=E/E_0$, or
$x=e^{E/E_0}$ for exponential self-similarty. Condition 3 leads in a super-position of the time dependent DF,
which forms the PL DF. The solution for $f(x)$  is derived here from the simulation, and in principle is set
by the system dynamics. 

The DF PL index $\Gamma$ is independent of the functional form
of $f(x)$, and is derived from two relations: I, the time dependence of the characteristic scale, say $E_0(t)$ in the simulations here, and
II, the boundary condition time dependence, say a constant particle injection rate. 
So, the derivation of the DF $\Gamma$ in a given system does not necessarily require a full solution
for the system dynamics, and only relations I + II may be enough to derive $\Gamma$.

Any physical system must have characteristic scales, say $E_{\rm initial}$ and $E_{\rm final}$, as it must start and end. This will inevitably set 
a maximal range of scales, say the range of $E$ values, over which the PL DF holds. The scale-free 
approximation requires $E\gg E_{\rm initial}$ and $E\ll E_{\rm final}$ (or the reverse if $E_0(t)$ goes down instead of up with $t$).
As a result, an observed PL DF must extend over a significant dynamical range, 
say two orders of magnitude or more, to minimize the edge effects at the distribution bundaries and have 
a significant dynamic range where the scale-free  self-similar solution holds to a good approximation.

Apart from the boundary value characteristic scales, the system
dynamics can change at some intermediate characteristic scale $E_{\rm char}$, say at $E_{\rm k}=mc^2$ where the dynamics
switches from non-relativistic to relativistic, which leads to a change in $\Gamma$. The PL solution will now hold on the separate scales
of $E_{\rm initial}\ll E_{\rm k}\ll mc^2$ and $mc^2\ll E_{\rm k}\ll E_{\rm final}$. 

We note in passing that conditions 1-3 
may be necessary for the formation of a PL DF, given Euler's theorem of a $1:1$ correspondence between self-similarity and a PL solution. If true, then the formation of a PL DF  
may not just be a possible solution
when conditions 1-3 apply, but rather the only solution. 

Why are PL DF so prevalent in nature? Condition 1 is prevalent in nature as the Universe is generally far
from statistical equilibrium. There was a Big Bang, everything is evolving, and in most systems we can tell if the time arrow is switched.
Condition 2 is likely common when the dynamics is simple, i.e. it is set by a small number of free parameters,
or equivalently a small number of characteristic scales. Condition 3 represents systems in which the boundary
conditions, which are set by the environment outside the system, evolve on longer time scales than the dynamical
timescale within the system. The boundary condition is then approximately constant in time, and the system can therefore reach a steady state (but not equilibrium) solution. 

The prevalence of conditions 1-3 may explain why PL DF are so common in nature. 
One may not need a full solution for the DF evolution in a given system, in order to derive 
the range where the PL solution holds and the PL index value. 

\section*{Data Availability}
The code that supports the findings of this study is available in Ref.~\cite{gitelman_2026_18642015}.


\appendix

\section{\label{apen:rel_freq} The relativistic collision rate}

The correct form of the equilibrium distribution in relativistic gases has been debated in recent years, with some proposals \cite{cubero2007} suggesting alternatives to the standard Jüttner distribution, such as the modified Jüttner form 
\cite {dunkel2009, Molnar2020}. 

In light of this debate, it is useful to highlight the covariant expression for the collision rate which, when properly implemented, ensures relaxation toward the Jüttner distribution. Following \cite{Landau1980Classical}, the number of binary collisions occurring in a spacetime volume element $dV\times dt$ is given by the Lorentz-invariant relation:
\begin{eqnarray}
    d\nu = \sigma_{\rm rel}\, n_1 n_2\, g_{12}\, dV dt,
    \label{eq:A1}
\end{eqnarray}
where
\begin{eqnarray}
    g_{12} = \sqrt{(\vec{\beta}_1 - \vec{\beta}_2)^2 - (\vec{\beta}_1 \times \vec{\beta}_2)^2}
\end{eqnarray}
is the relative velocity factor, and $\vec{\beta}_{1,2}$ are particles velocity (in units where $c=1$), $n_{1,2}$ are their number densities, and $\sigma_{\rm rel}$ is the cross section in the rest frame of one of the particles. This form guarantees that the collision rate transforms covariantly under Lorentz transformations.

In 1-D kinematics, Eq.~\ref{eq:A1} reduces to $d\nu = \sigma_{\rm rel} |\beta_1 - \beta_2| n_1 n_2\, dV\, dt$, and indeed simulations in one dimension (e.g., \cite{cubero2007}) recover the Jüttner distribution. However, this 1-D result does not generalize to 3-D, as the full invariance structure is lost in reduced dimensionality.

Monte Carlo simulations in 3-D by \cite{peano2009} explicitly demonstrate that incorrect sampling—e.g., uniform selection of pairs with constant acceptance—leads to systematic deviations from the Jüttner form. Only when the collision probability is weighted by the covariant flux factor $g_{12}$ does the system evolves to the standard Jüttner equilibrium. In our simulations, although pairs are drawn uniformly, we accept collisions with probability proportional to $g_{12}$, thereby enforcing the correct effective rate. 

Thus, in the context of ongoing discussion regarding the relativistic equilibrium distribution, Eq.~\ref{eq:A1} plays a central role: it provides the kinetic foundation from which the Jüttner distribution naturally emerges when collisions are modeled consistently with relativistic invariance.

\section{\label{apen:mean_e_pec} Equipartition, equal $\langle E_{\rm k}\rangle$ or equal $T$?}

A system of heavy and light point particles in statistical equilibrium is expected to have reached equipartition, 
i.e. equal DF of the particles $E_{\rm k}$. The DF is defined uniquely by a single parameter $T$. However, the
DF can also be defined by its $\langle E_{\rm k}\rangle$, and therefore there is a 1:1 correspondence between $T$
and $\langle E_{\rm k}\rangle$. However, the equivalence of $T$ and
$\langle E_{\rm k}\rangle$ holds only when both particle populations follow the same DF.
In the MB limit $\langle E_{\rm k}\rangle=3/2\times k_{\rm B}T$, while in MJ $\langle E_{\rm k}\rangle=3\times k_{\rm B}T$
(eq.~\ref{eq:MeanEnergy}).
What happens when we have a mix of relativstic and non relativistic particles in statistical equilibrium,
will their DF have the same $T$ or the same $\langle E_{\rm k}\rangle$?

For example, in plasma at $T=10^{11}$~K, where the characteristic kinetic energy $kT\simeq 10$~MeV, 
the electrons are relativistic with a mean $\gamma \simeq 20$, while the protons are non-relativistic, 
with a mean $\beta\simeq 0.15$. That is the heavy and slow protons follows the MB DF, while the light and fast 
electrons follow MJ (of course MJ holds at all energies, but below we use MJ for the relativistic limit).
The MB and MJ DF have a different functional form, so clearly equipartition cannot lead in this case to identical
DF for the two populations. What will then be the relation between the DF of the two populations 
when the system reaches equipartition?

\begin{figure}
\includegraphics[width=8 cm]{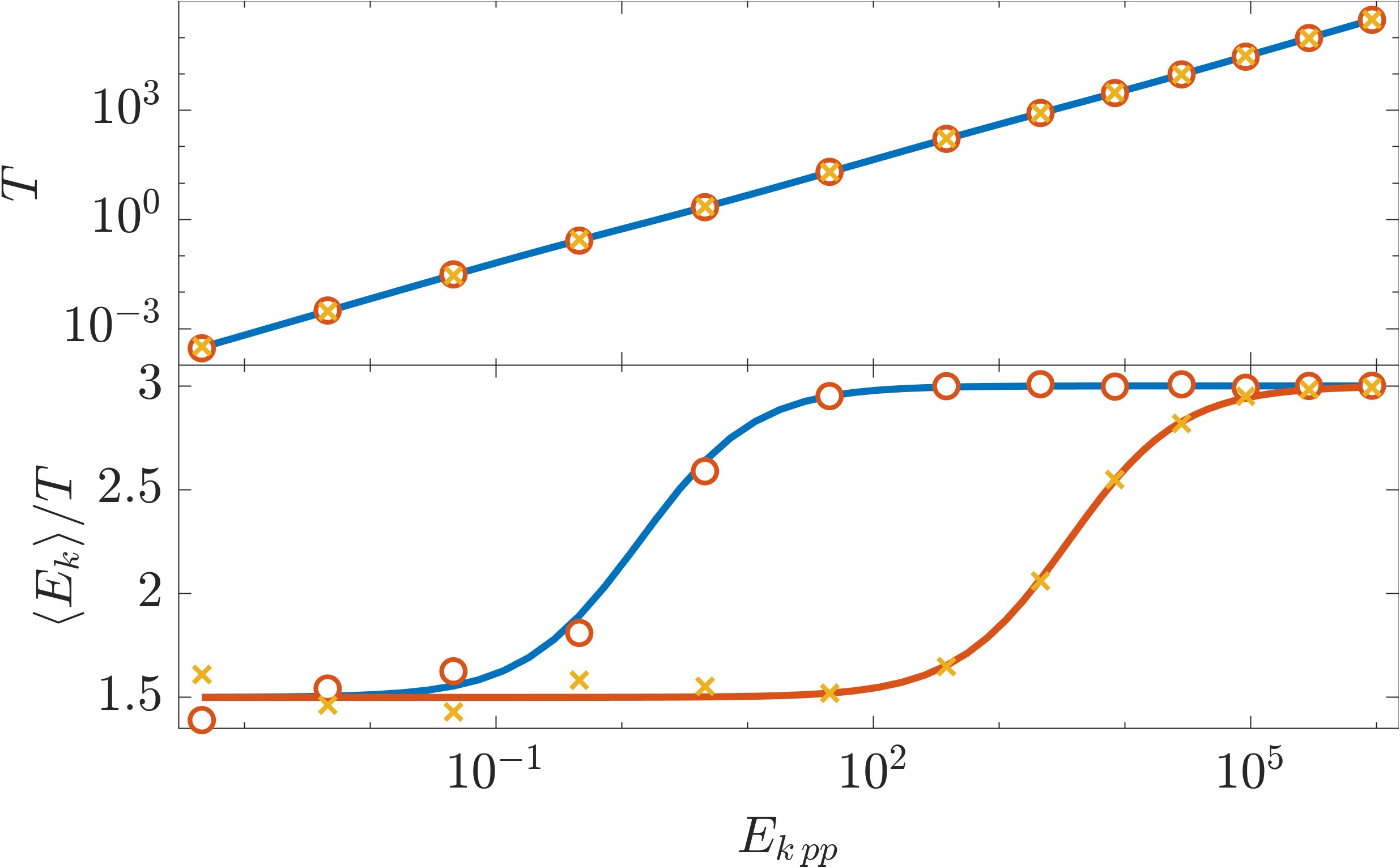}
\caption{\label{fig:Ratio_mean_e} The equilibrium solutions for a system composed of light ($m_1=1$)
and heavy ($m_2=1836$) particles with a range of mean
kinetic energy per particles $E_{\rm k,pp}$. The x-axis denotes $E_{\rm k,pp}$ 
in each simulation. The derived DF for the two populations is fit with a MJ DF in each simulation. 
{\bf Upper panel} The best fit $T$ for each population in each simulation. 
The solid lines presents the expected relations for the light particles (blue line) and heavy
particles (red line). {\bf Lower panel} As in the upper panel, for the measured 
$\langle E_{\rm k}\rangle/T$. In the non-relativistic limit  ($E_{\rm k,pp}\ll 1$) 
both populations follow the same MB DF, as expected for a system in equipartition,
leading to identical $T$ and $E_{\rm k,pp}\ll 1$ in each simulation.
Similarly, in the ultra relativsitic limit ($E_{\rm k,pp}\gg 10^3$),
both populations follow the same MJ DF. 
In the intermediate regime, the heavy particles follow
MB and the light particles follow MJ. Both populations have the same $T$, and 
therefore different $\langle E_{\rm k}\rangle$. Thus, equipartition of point particles
always imply equal $T$, but not necessarily equal $\langle E_{\rm k}\rangle$.
}
\end{figure}

Figure~\ref{fig:Ratio_mean_e} presents the results of a simulation to address the question above.
The simulation is for a system with two types of particles, light
particles with $m_1=1$ and heavy particles with $m_2=1836$, equivalent to an electron + proton system at low
densities, where the electrostatic forces are mostly negligible. In each simulation all particles start
with a given initial value of $\gamma\beta$, that is the system is initially far from equilibrium. The simulation 
extends in time until the system
reaches a statistical equilibrium. We then derive the DF of $E_{\rm k}$ for the light and for the heavy
particles, and derive $T$ from the best analytic MJ DF. The upper panel presents the derived $T$ for the two
populations as a function of the average initial energy per particle $E_{\rm{k,pp}}$. Clearly, the DF of
the two populations are characterized by the same $T$. This result is not surprising at $E_{\rm{k,pp}}<0.1$
and at $E_{\rm{k,pp}}>10^4$, where both populations are non relativistic or relativistic, and therefore 
characterized by the same DF. In these limits the equipartition leads to equal DF in the two populations,
and therefore equal $T$ values. However, the same applies in the intermediate $T$ regime where the DF are
different.

Figure~\ref{fig:Ratio_mean_e} lower panel, presents the derived $\langle E_{\rm k}\rangle/T$ (for $k_{\rm B}=c=1$)
in the two populations as a function of $E_{\rm{k,pp}}$ for each simulation. Clearly, in the intermediate
regime $0.1<E_{\rm{k,pp}}<10^4$ the two population have different $\langle E_{\rm k}\rangle$. The mean
kinetic energy of the light particles which are relativistic can reach twice the kinetic energy of the heavy particles
which are non relativistic, as expected if both DF have the same $T$ values, and  $\langle E_{\rm k}\rangle/T$
drops from 3 to 3/2 in the transition from the relativistic to the non relativistic limits.

\section{\label{chap:PL_to_MB_kappa_DF} Transition from a Power-Law to a Maxwellian DF: does a $\kappa$ DF forms?}

\begin{figure*}
\begin{center}

\includegraphics[width=0.4 \textwidth]{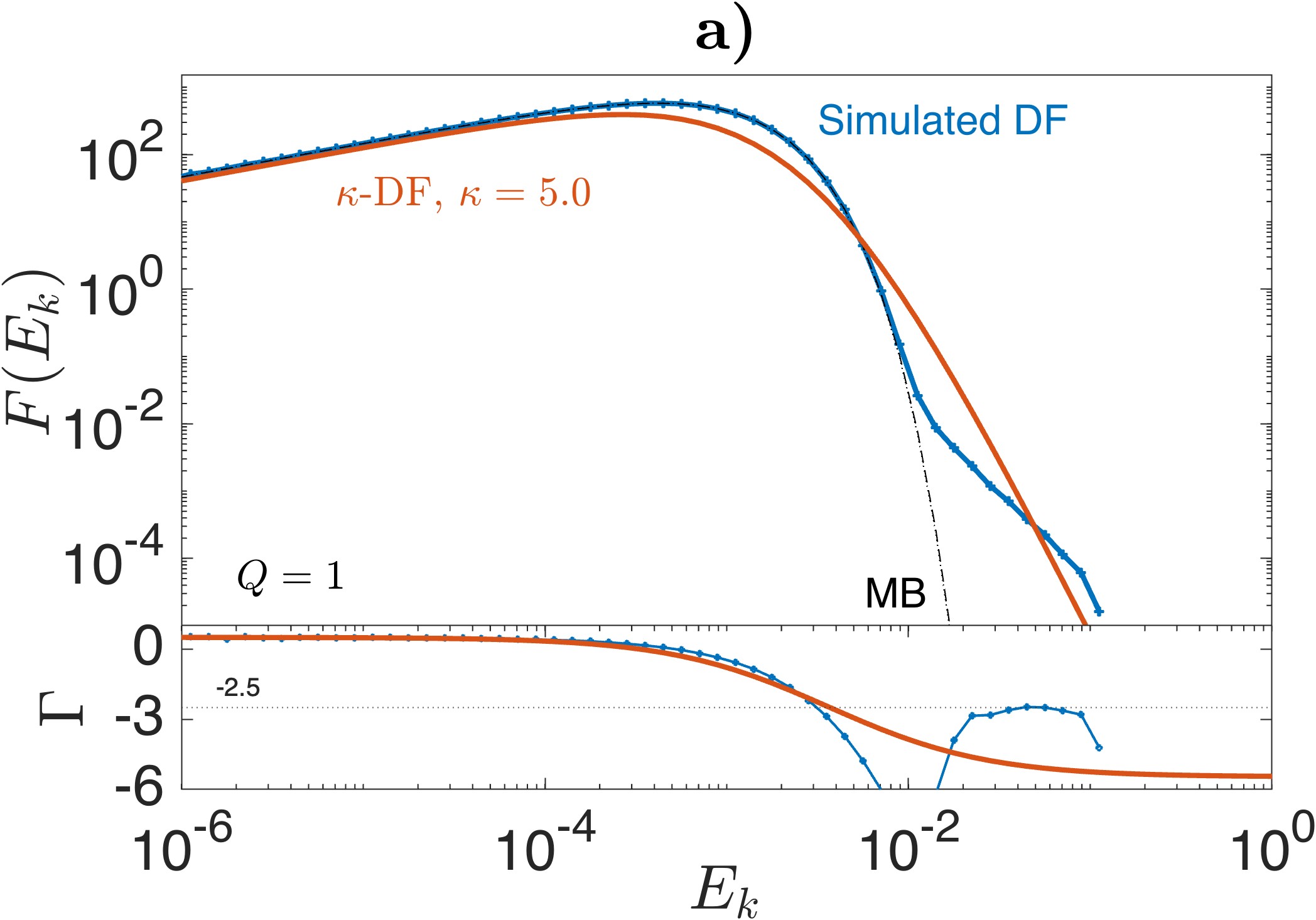}
\includegraphics[width=0.4 \textwidth]{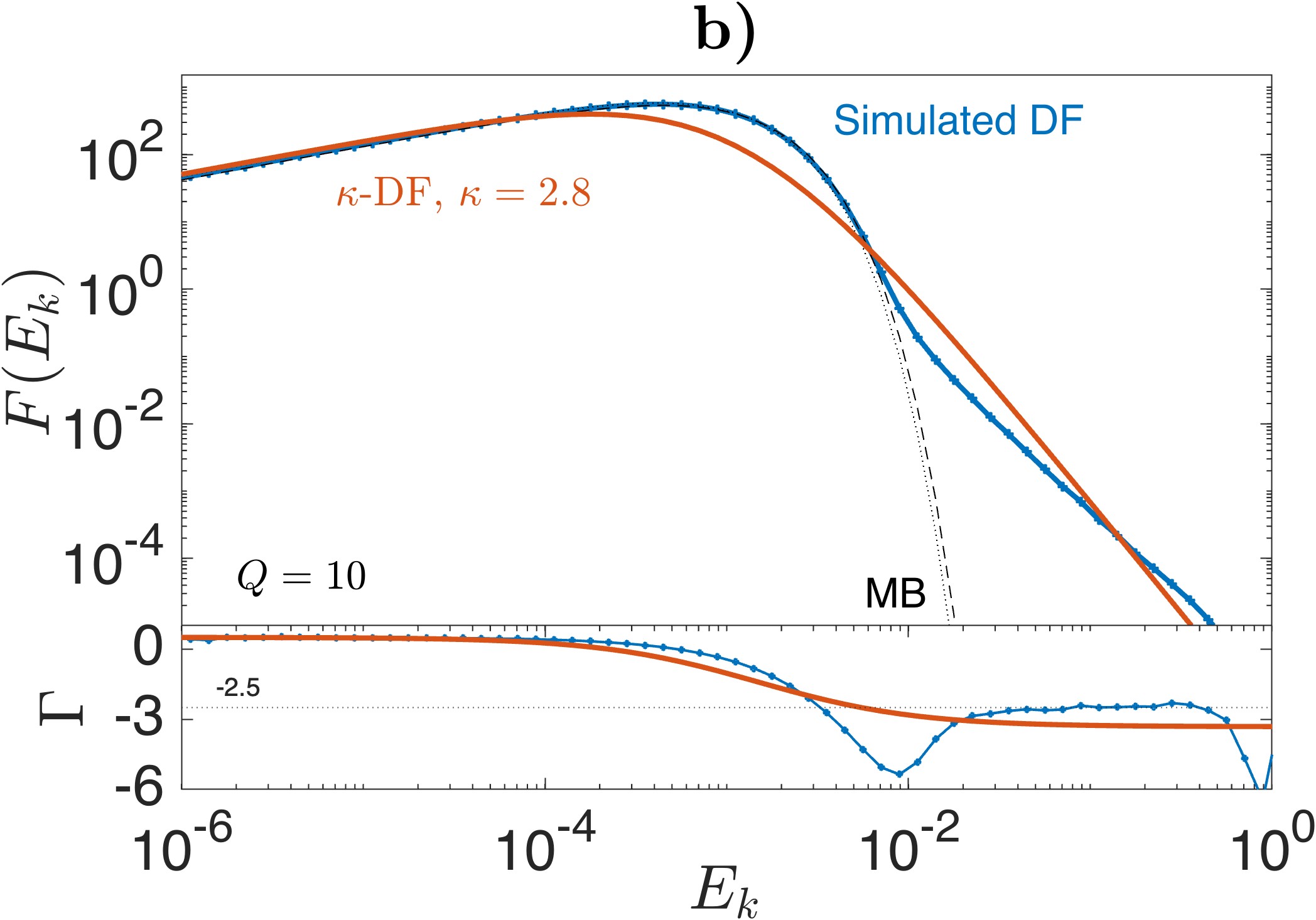}
\includegraphics[width=0.4 \textwidth]{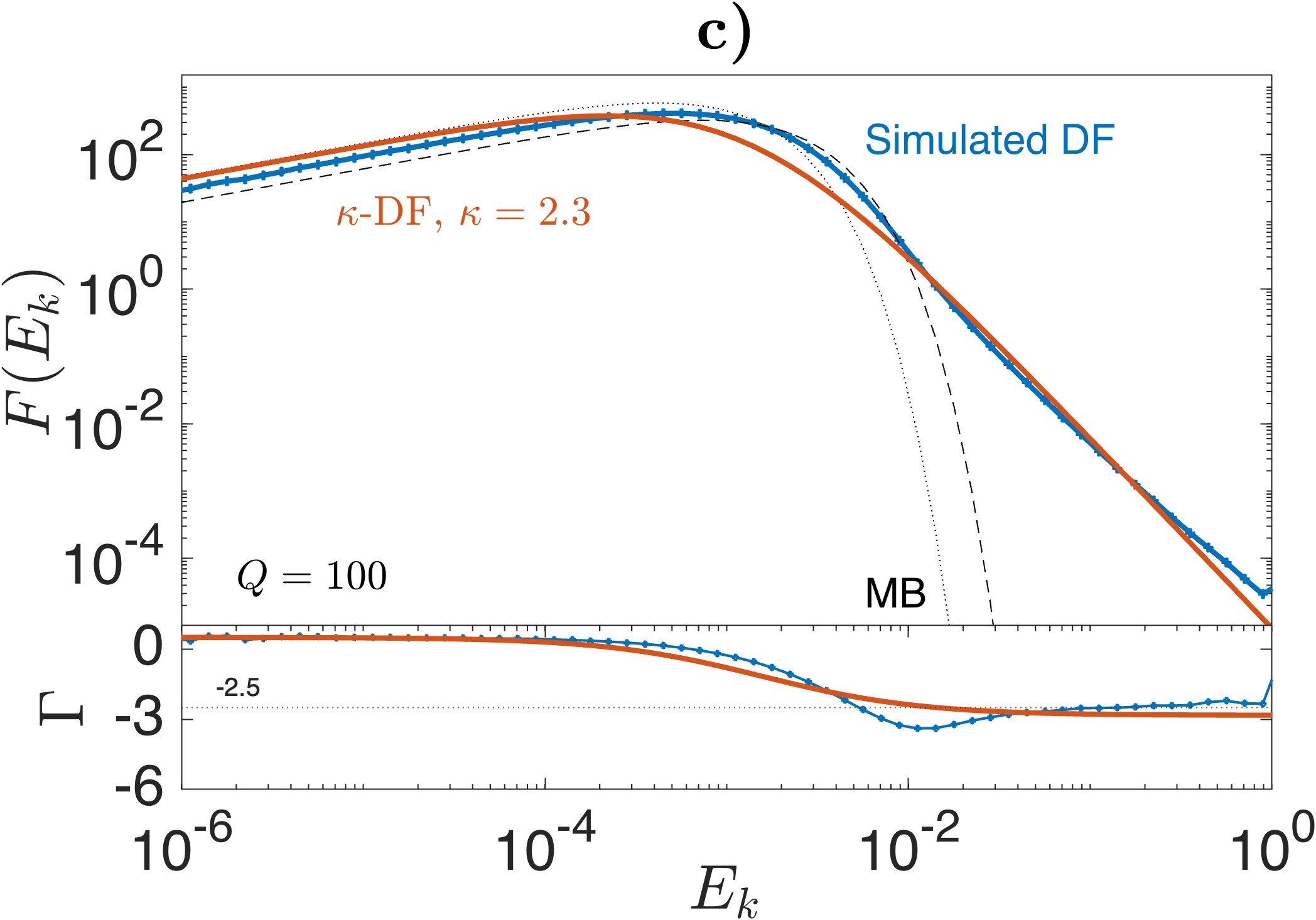}
\includegraphics[width=0.4 \textwidth]{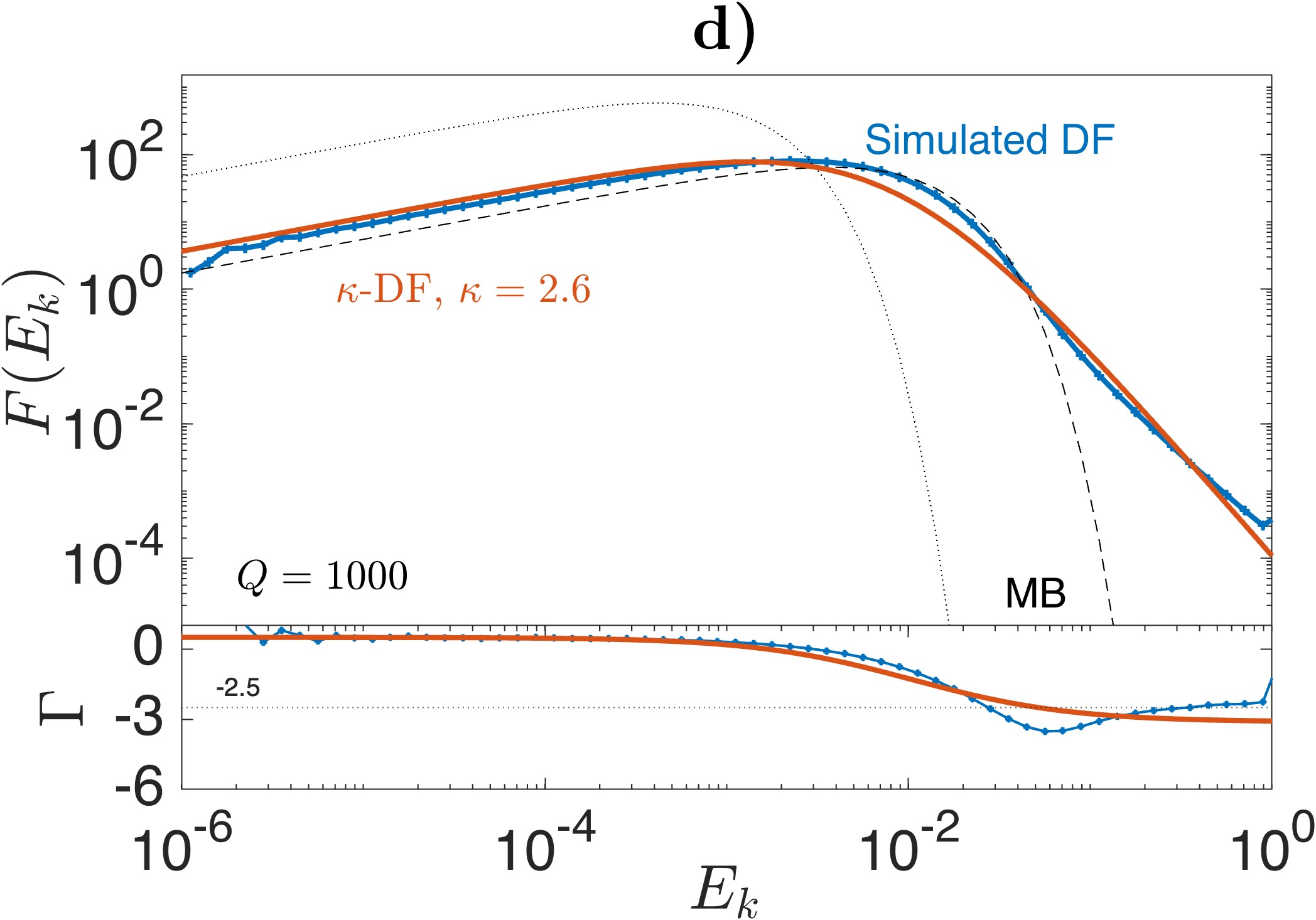}

\caption{\label{fig:one_pop_diff_rel} Does a cascade simulation produces a $\kappa$ DF? The figure shows the distribution functions at $t_f = 100$ for varying energy injection rates $Q$ of particles with $E_k=1$. The simulation result is shown in blue; the initial MB is plotted as a dotted line and the final MB as a dashed line. The best-fit $\kappa$ DF is in red-orange. At low injection rates ($Q = 1$, $10$; panels a–b), the ratio $E_{\rm inj}/E_{\rm initial} =0.01, 0.1$, and the DF retains the thermal MB core with a power-law nonrelativistic cascade tail of index $\Gamma \simeq -2.5$. At higher rates ($Q = 100$, $1000$; panels c–d), where $E_{\rm inj}/E_{\rm initial} = 1, 10$, the thermal core shifts to higher energies, but remains undistorted in shape. The DF remains MB + PL, and the PL $\Gamma$ remains unchanged. These results suggest that the DF in thermal plasmas with high-energy particle injection is a two-component structure, MB + PL. The formation of a $\kappa$ like DF requires fine tuning of the injection rate.}  

\end{center}
\end{figure*}

In the cascade simulations above we assumed a constant (infinitely large) thermal background.
Below we explore the effect of the PL formation mechanism on the background thermal MB DF for a finite size background. 
The initially thermal background is also followed in the simulation, including scattering among the background particles.
The simulations includes energy exchange between the PL and the thermal components which becomes comparable 
to the initial thermal energy. The question we aim
to answer is whether the DF of the combined system is well approximated by a simple superposition of a MB and a PL DFs,
or whether the end result is a significant modification of the MB DF,  in particular whether it transforms into the $\kappa$ DF.

The  $\kappa$ DF is widely used to model plasmas with both thermal and suprathermal components. For example, the solar wind and planetary magnetospheres \citep[e.g.][]{Pierrard2010, Livadiotis_2011,Livadiotis_2013}, as well as suprathermal particles in collisionless shock environments \citep{ARBUTINA2021}. 

The $\kappa$ DF is given by:
\begin{eqnarray}
    F(E_k)\propto \frac{\sqrt{E_k}}{\left(1+\frac{E_k}{\left(\kappa - \frac{3}{2}\right) T}\right)^{\kappa + 1}},
\end{eqnarray}
with $T = \frac{2}{3} \langle E_k \rangle$. It follows a MB DF $F(E_k)\propto E_k^{1/2}$ below the DF peak energy, and transitions smoothly to a PL with an index $F(E_k)\propto E_k^{-\kappa-1/2}$ at $E_k\gg T$. The  $\kappa$ DF has also been proposed as a possible explanation for some temperature diagnostics in H II regions and planetary nebulae \citep{Nicholls2012}, though their applicability in such photoionized environments has been questioned, with thermalization arguments suggesting that $\kappa$ distributions are unlikely to form \citep{Draine2018}. We use our simulations to explore whether the $\kappa$ DF naturally describes
a system which includes a PL DF formed by cascade collisions on a MB component.

Below we present a cascade simulation which follows both the formation of the PL DF, and the time evolution of the
thermal background MB DF, and explores whether a $\kappa$-like DF can form in such a system. 
Each simulation includes $N = 10^7$ particles (some simulations are repeated many times and the final results summed to 
reduce the poisson noise) with $m = 1$ and initial momentum $\gamma \beta = 0.05$ in random directions. At each time interval $\Delta t$, the kinetic energy of a random set of $n = 10$ particles is increased by a fixed amount $e_i = 1$, which corresponds to an energy injection rate of $Q = n e_i/\Delta t$.

The system relaxes to a MB  (see Section~\ref{subsec:valid}) on the thermalization time $t_{\rm th} \sim 10$, 
based on our earlier simulations (Figure~\ref{fig:code_val}) where we get 
$t_{\rm th} \sim 5$ for $\beta \sim \gamma \beta = 0.1$, or equivalently after roughly one collision per particle. The injection interval $\Delta t$ in the simulations is varied between $10^{-2}$ and $1$, always shorter than $t_{\rm th}$, to avoid complete
thermalization of the PL component.

The simulations include only heating by the injected particles, and therefore cannot reach a steady state.
All simulations are therefore stopped at $t_f = 100$. 
The total energy injected during the simulation time is
\[
E_{\rm inj} = t_f \cdot Q = t_f \cdot \frac{n e_i}{\Delta t}.
\]
This is compared with the total initial kinetic energy in the system,
\[
E_{\rm initial} = N \cdot \left( \sqrt{1 + (\gamma \beta)^2} - 1 \right) \cdot m,
\]
which for $\gamma \beta = 0.05$ $m = 1$ and $N = 10^7$, gives $E_{\rm initial} \approx 1.25\times 10^4$. The ratio $E_{\rm inj} / E_{\rm initial}$ characterizes the energy loading of the system. We simulate injection rates in the range $Q = 1 - 10^{3}$, which 
correspond to $E_{\rm inj} / E_{\rm initial}\sim 10^{-2}- 10$, i.e. minimal to possibly major changes in the initial MB DF.

Figure~\ref{fig:one_pop_diff_rel} presents the results. Panels a and b present the low injection rate $Q=1$, $10$, 
where $E_{\rm inj} / E_{\rm initial} = 0.01, 0.1$. The MB DF remains unchanged, with a sharp transition to a PL DF. The PL index
is $\Gamma=-2.5$, as derived above for the non relativistic cascade simulation for a constant MB background
(Figure ~\ref{fig:cascade_evol_rel}).
The PL normalization increases by an order of magnitude, when $Q$ increases from 1 to 10. The transition $E_k$
is sharp, as indicated by the deep in the $\Gamma$ plot, and occurs where the PL and the MB amplitudes match. 

The best fit $\kappa$ DF matches the MB DF at $E_k\ll kT$, where both scale as $E_k^{1/2}$. However, above the 
DF peak the $\kappa$ DF becomes significantly steeper than the $\Gamma=-2.5$ of PL derived in the simulation. 
Also, the $\kappa$ DF shows a smooth transition in $\Gamma$ with $E_k$, without the sharp deep the simulations give.
The $\kappa$ DF corresponds to a PL with $\Gamma=0.5$ which transforms smoothly to a PL with $\Gamma=-\kappa-0.5$,
while the simulations give a MB DF which changes sharply to a $\Gamma=-2.5$ PL above some $E_k$, which depends
on the PL component normalization. 

Figure~\ref{fig:one_pop_diff_rel}, panel c, presents the results for $Q=100$, where ratio $E_{\rm inj} / E_{\rm initial} \sim 1$.
The simulation results are smilar, a MB DF which transforms to a PL DF which increases in amplitude by a further
factor of 10. The MB DF is also affected, and is shifted to a somewhat higher temperature. Since the PL intersects 
the MB closer to the peak, where the local index is less steep, the deep in the $\Gamma$ plot is shallower.
The offset of the best fit $\kappa$ DF from the simulation is now smaller, with an asymptotic $\Gamma=-2.8$ which
is closer to the simulation result $\Gamma=-2.5$. Panel d for  $Q=1000$ and $E_{\rm inj} / E_{\rm initial} \sim 10$, shows
that despite the major energy injection the DF remains composed of a MB with a sharp transition to a PL.
The PL normalization increases again by a factor of 10, but now the MB also shifts by a factor of $\sim 10$,
so the simulation $\Gamma$ remains similar to the previous case, and so is the $\kappa$ DF.

The simulations show that injecting particles drawn from a MB to high $E_k$ ($\gg kT$) leads to the formation of a PL DF
with a sharp transition to a MB with no distortions close to the transition region. The amplitude of the PL
is set by the particle injection rate, and its index is set by the cascade dynamics. The derived MB + PL DF 
is generally different from the $\kappa$ DF, where the PL normalization is not a free parameter. 

We therefore suspect that plasma with an evidence for suprathermal particle tail formed by the injection of high
energy particles, may be more accurately 
described by a MB + PL DF, rather than by the $\kappa$ DF \cite[see also][]{Draine2018}. 

\bibliography{PLtoMB}
\end{document}